\documentclass[10pt,openany,a4paper]{article}
\usepackage{geometry,color,framed}
\usepackage{amsmath,amssymb,mathtools,slashed,xcolor,mdframed}
\usepackage[vcentermath,enableskew]{youngtab}
\usepackage{cancel, pifont,multirow}
\usepackage{tikz}
\usetikzlibrary{positioning}
\usetikzlibrary{decorations.text}
\usepackage{amstext}
\usepackage{graphicx}
\usepackage[section]{placeins}
\usepackage{verbatim}
\usepackage{bm}
\usepackage{caption, subcaption, float}
\usepackage[colorlinks=true, 
            linkcolor=blue, citecolor=blue,
            menucolor = red,
            urlcolor=blue,linktoc=all]{hyperref}
\makeatletter

\newcommand{\beq}{\begin{eqnarray}}
\newcommand{\eeq}{\end{eqnarray}}
\newcommand{\non}{\nonumber\\}

\newcommand{\mc}{\mathcal}
\newcommand{\mb}{\mathbb}
\newcommand{\mf}{\mathbf}

\newcommand{\ben}{\begin{enumerate}}
\newcommand{\bei}{\begin{itemize}}
\newcommand{\eni}{\end{itemize}}
\newcommand{\enn}{\end{enumerate}}
\newcommand{\ra}{\rightarrow}
\newcommand{\rr}{\right}
\newcommand{\lf}{\left}

\newcommand{\xdownarrow}[1]{%
  {\left\downarrow\vbox to #1{}\right.\kern-\nulldelimiterspace}
}

\def\a{{\alpha}}

\def\l{{\lambda}}

\def\s{{\sigma}}

\usepackage{tikz}
\usetikzlibrary{matrix,shapes,arrows,positioning,chains}
\tikzset{
	 connector/.style={
        -latex,
        font=\scriptsize
    },
    rectangle connector/.style={
        connector,
        to path={(\tikztostart) -- ++(#1,0pt) \tikztonodes |- (\tikztotarget) },
        pos=0.5
    },
    rectangle connector/.default=-2cm,
    straight connector/.style={
        connector,
        to path=--(\tikztotarget) \tikztonodes
    }
}
\makeatother
\title{Hyperbolic recurrent neural network \\ as the first type of non-Euclidean neural quantum state ansatz}
\author{H. L. Dao\footnote{espoirdujour1162@gmail.com}}
\date{\today}
\date{\today}
\begin{document} 
\maketitle
\begin{abstract}
In this work, we introduce the first type of non-Euclidean neural quantum state (NQS) ansatz, in the form of the hyperbolic GRU (a variant of recurrent neural networks (RNNs)), to be used in the Variational Monte Carlo method of approximating the ground state wavefunction for quantum many-body systems. In particular, we examine the performances of NQS ansatzes constructed from both conventional or Euclidean RNN/GRU and from hyperbolic GRU in the prototypical settings of the one- and two-dimensional transverse field Ising models (TFIM) and the one-dimensional Heisenberg $J_1J_2$ and $J_1J_2J_3$ systems with varying values of $(J_1,J_2)$ and $(J_1, J_2, J_3)$. By virtue of the fact that, for all of the experiments performed in this work, hyperbolic GRU can yield performances comparable to or better than Euclidean RNNs, which have been extensively studied in these settings in the literature, our work is a proof-of-concept for the viability of hyperbolic GRU as the first type of non-Euclidean NQS ansatz for quantum many-body systems. Furthermore, in settings where the Hamiltonian displays a clear hierarchical interaction structure, such as the 1D Heisenberg $J_1J_2$ \& $J_1J_2J_3$ systems with the first, second and even third nearest neighbor interactions, our results show that hyperbolic GRU definitively outperforms its Euclidean version in almost all instances. Additionally, in the 2D TFIM setting, 1D hyperbolic GRU NQS also outperforms 1D Euclidean GRU NQS, due to the presence of a hierarchical interaction structure comprising the first and $N^{th}$ nearest neighbor interactions, introduced by the rearrangement of the 1D spin chain to mimic the 2D $N\times N$ lattice. The fact that these results are reminiscent of the established ones from natural language processing where hyperbolic RNNs almost always outperform Euclidean RNNs when the training data exhibit a tree-like or hierarchical structure leads us to hypothesize that hyperbolic GRU NQS ansatz would likely outperform its Euclidean counterpart in quantum spin systems that involve different degrees of nearest neighbor interactions. Finally, with this work, we hope to initiate future studies of other types of non-Euclidean neural quantum states beyond hyperbolic GRU.
\end{abstract}

\tableofcontents
\section{Introduction}
With the advance of deep learning in physics, neural networks have been established as efficient variational trial wavefunctions (variational ansatzes) to  estimate the ground states of quantum systems starting with the work of \cite{1606-rbm} that introduced restricted Boltzmann machines (RBM) as viable candidates for the trial wavefunctions in the spin model settings of 1D Ising model, 1D and 2D Heisenberg models. Since then, in addition to RBM being further employed as variational ansatz in various quantum systems \cite{1610-rbm}, \cite{1905-rbm}, many other types of neural networks have also been constructed and shown to be viable trial/variational wavefunctions in a variety of different problem settings in quantum and condensed matter physics ranging from spin systems \cite{1606-rbm} - \cite{rnn-25}, fermion systems \cite{fer-1}-\cite{fer-4}, itinerant boson systems \cite{itinerant-boson-1}, \cite{itinerant-boson-2} to matrix models \cite{prx_22}, \cite{2409-bnaf} in the recent literature. Some notable examples of these include ANN-based ansatzes \cite{1704-ann}, \cite{1709-ann}, CNN-based ansatzes\footnote{ANN stands for Artifical Neural Network, while CNN stands for Convolutional Neural Network.} \cite{cnn-1807}, \cite{cnn-1903}, \cite{gcnn-2211}, RNN-based ansatzes \cite{rnn_20}, \cite{rnn_22}, \cite{rnn-24}, \cite{rnn-25}, transformer-based ansatzes \cite{2211-transformer}, \cite{transformer-2311} \cite{2306-transformer}, \cite{2406-transformer}, \cite{transformer-25} and even generative networks like autoregressive \cite{1902-autoregressive}  and normalizing-flow-based ansatzes \cite{prx_22}, \cite{2409-bnaf}. Collectively, this class of trial wavefunctions based on neural network models are known as neural quantum states (NQS). Very recently, a type of foundation NQS, based on the transformer architecture and introduced in the work \cite{fnqs-2502}, is the first type of versatile NQS capable of processing multimodal inputs comprised of spin configurations and Hamiltonian couplings, and generalizing beyond those Hamiltonians encountered during its training phase. Beyond the task of approximating the ground state, extensions of NQS for approximating the excited states are also an active direction of research \cite{excited-1}. A recent review of NQS, including the different neural network architectures used and their many applications in quantum many-body physics can be found in \cite{2402-nqs-review} and references therein.
\\\\
In this work, we present the first type of non-Euclidean NQS based on hyperbolic recurrent neural networks (RNNs). This type of hyperbolic RNNs, whose underlying hyperbolic geometry is that of a Poincar\'e ball,  was introduced in the work of \cite{ganea-1805} in the context of natural language processing for the tasks of sentence embeddings, textual entailment and noisy-prefix recognition. For these tasks, hyperbolic RNNs have been shown to outperform their Euclidean counterparts, especially in settings when there the training data possess an inherent hierarchical structure \cite{ganea-1805}, \cite{hyprnn-review-2101}, \cite{hyprnn-lorentz-1806}, \cite{2010-hyprnn}. Besides the Poincar\'e model, another variant of hyperbolic RNNs based on the Lorentz model \cite{hyprnn-lorentz-1806} was proposed and deployed to carry out the NLP task of learning hierarchies in unstructured similarity scores.
Besides hyperbolic RNNs, other types of hyperbolic networks, such as hyperbolic CNNs \cite{2006-hypcnn}, hyperbolic attention networks \cite{1805-hyp-attn}, hyperbolic transformers \cite{hyprnn-lorentz-2105}, hyperbolic normalizing flows \cite{hyprnn-review-2101}, also have been introduced in quick succession to tackle different NLP-related  tasks in deep learning. In \cite{hyprnn-lorentz-2105}, the authors demonstrated the superior performances of Lorentz-based hyperbolic networks, including hyperbolic transformers, over their Euclidean versions, in the tasks of machine translation and knowledge graph completion. Regardless of the underlying hyperbolic model, i.e. either Poincar\'e or Lorentz, hyperbolic neural networks have consistently demonstrated their ability to better learn from and model data with latent hierarchies \cite{ganea-1805}-\cite{2010-hyprnn}. Beyond hyperbolic geometry, the possibility of constructing and deploying other neural networks based on different types of non-Euclidean geometries in deep learning has opened up new and exciting research directions for the relatively new field of non-Euclidean machine learning, whose examples include mixed-curvature variational autoencoders \cite{1901-vae}, constant curvature graph convolution neural networks \cite{1911-cnn} among others. 
\\\\
In the aforementioned tasks involving hierarchical, tree-like data structures, the obvious outperformances delivered by hyperbolic neural networks when benchmarked against their Euclidean counterparts can be largely attributed to the hyperbolic geometry underlying the construction of these networks. It was known from the works of \cite{combinatorica-95}, \cite{sarkar-11}, \cite{krioukov-09}, \cite{krioukov-10} and \cite{sala-2018} that hyperbolic space has the capacity to embed or represent tree structures with very low distortion and arbitrarily small errors. This is because in $D$-dimensional hyperbolic space $\mathbb{H}_D$, the volume grows exponentially with respect to the radius, while in  $D$-dimensional Euclidean space $\mathbb{E}_D$, the volume only grows polynomially\footnote{To be concrete, let's consider the case of 2D hyperbolic space $\mathbb{H}_2$ and 2D Euclidean space $\mathbb{E}_2$. In $\mathbb{H}_2$ with the negative constant curvature $-\zeta$ where $\zeta>0$, the length of the circle and the area of the disk of radius $R$ are $2\pi \sinh(\zeta R)$ and $2\pi\,(\cosh\zeta R-1)$, respectively. Both of these quantities grow as $e^{\zeta R}$, i.e. exponentially with respect to $R$. In Euclidean space $\mathbb{E}_2$, the circle length and disk area of radius $R$ are $2\pi R$ and $\pi R^2$, respectively. These only grow polynomially with respect to $R$. }. Due to the exponential expansion of space in hyperbolic geometry, there exists a direct link or map between a $b$-ary tree (a tree with branching factor $b$) and hyperbolic space, with the former informally thought of as a discrete version of the latter\footnote{This is the case because the number of nodes at a distance $R$ away from the root of a $b$-ary tree is $(b+1)b^R$, which grows as $b^R$ as $R$ increases, similar to the circle length in hyperbolic space}. As such, hyperbolic spaces allow for almost isometric embeddings of tree structures\footnote{In particular, the work \cite{sarkar-11} showed that tree structures can be embedded without distortion in 2D hyperbolic space, a Poincar\'e disk, while the work \cite{sala-2018} extended this embedding to $D$-dimensional hyperbolic space.}, which require an exponential amount of space for branching, while Euclidean spaces (of any dimensions) do not \cite{combinatorica-95}, \cite{sarkar-11}, \cite{krioukov-10}, \cite{sala-2018}.
\\\\
Here, in the context of quantum many-body physics, we will reconfigure the hyperbolic GRU (Gated Recurrent Unit) as defined in \cite{ganea-1805} to be used as the first type of non-Euclidean NQS ansatz in the prototypical settings of the 1D and 2D transverse field Ising model (TFIM) as well as the 1D Heisenberg $J_1J_2$ and $J_1J_2J_3$ models. These models are the most basic testing grounds for a variety of experiments involving neural quantum states as demonstrated in the earliest works \cite{1606-rbm}, \cite{1704-ann} on this topic. With the aim of determining the viability and competitiveness of the hyperbolic GRU as an NQS ansatz, in each of these settings (Ising and Heisenberg models), the performance of the hyperbolic GRU ansatz will be benchmarked firstly against the DMRG (Density Matrix Renormalization Group) results and secondly against that of the conventional or Euclidean RNN/GRU ansatz introduced in the work of \cite{rnn_20}. Our obtained results not only demonstrate this feasibility but also potentially open up the possibility of better performances with hyperbolic GRU in quantum settings that exhibit hierarchical structures. Although more work are needed to conclusively confirm this observation in quantum systems beyond those studied in this work, given these first promising results, we hope to highlight the potential of non-Euclidean NQS in quantum many-body physics.
\\\\
This paper is organized as follows. In section \ref{sec-vmc}, we recall the details of the Variational Monte Carlo (VMC) method to approximate the ground state energy of a quantum system. In section \ref{sec-hyprnn}, we summarize the mathematical details of the Poincar\'e model of hyperbolic space in \ref{sec-poincare} and describe the mathematical constructions of hyperbolic RNN (in \ref{sec-hrnn}) and of hyperbolic GRU (in \ref{sec-hgru}) from their respective Euclidean versions. In section \ref{sec-nqs-rnn}, we describe the construction of the neural quantum state trial wavefunction using RNNs, with the real case described in \ref{sec-wf-r} and complex case described in \ref{sec-wf-c}.  In section \ref{sec-res}, we present the main results obtained by using hyperbolic and Euclidean RNNs  as variational ansatzes in the settings of 1D TFIM (section \ref{sec-res-tfim}), 2D TFIM (section \ref{sec-res-2dtfim}), 1D Heisenberg $J_1J_2$ model (section \ref{sec-res-j1j2}) and 1D Heisenberg $J_1J_2J_3$ model (section \ref{sec-res-j1j2j3}). Section \ref{sec-concl} closes the paper with some concluding remarks. In the appendix, we include the convergence curves for the mean energy, obtained during the training process, of all VMC experiments. 
\\\\
The Python codes  used in this work to construct the hyperbolic GRU (using \texttt{Tensorflow})\footnote{For the construction of the hyperbolic GRU neural network, we adapted the Tensorflow-v1 (TF1) codes of \cite{ganea-1805} written for hyperbolic RNNs in the context of natural language processing tasks, which can be found at \href{https://github.com/dalab/hyperbolic_nn}{https://github.com/dalab/hyperbolic\_nn}, to the Tensorflow version 2 (TF2) used in this work} and to run all VMC experiments\footnote{For the VMC experiments as well as the construction of general RNN-based NQS ansatzes, we adapted the Tensorflow v1 (TF1) codes of \cite{rnn_20} found at \href{https://github.com/mhibatallah/RNNWavefunctions/}{https://github.com/mhibatallah/RNNWavefunctions/} to Tensorflow v2 (TF2) with the addition of the hyperbolic GRU ansatz. } as well as all the trained neural networks can be found at this Github repository: \href{https://github.com/lorrespz/nqs_hyperbolic_rnn}{https://github.com/lorrespz/nqs\_hyperbolic\_rnn}.
\clearpage
\section{Variational Monte Carlo (VMC)}\label{sec-vmc}
The Variational Monte Carlo (VMC) method, often used to train NQS, involves the process of sampling from a probability distribution represented by the square of the trial wavefunction/the variational ansatz and subsequently using these generated samples to calculate some obervables such as the ground state energy. 
In this section, we recall the derivation of the local energy formula in the Variational Monte Carlo (VMC) method \cite{mc-textbook}.
\\\\
Given a quantum Hamiltonian $H$ and the complete basis $|x\rangle$ satisfying $\sum_{|x\rangle} |x\rangle \langle x| = 1$, the trial/variational wavefunction $|\Psi\rangle$ of this quantum system represented by $H$ can be written in terms of the basis $|x\rangle$ as 
\beq
|\Psi\rangle = \sum_{|x\rangle}|x\rangle\langle x |\Psi\rangle  = \sum_{|x\rangle} \Psi(x) |x\rangle,
\eeq
where $\Psi(x) = \langle x|\Psi\rangle$. 
The variational energy is the expectation of the Hamiltonian with respect to $|\Psi\rangle$
\beq
E = \frac{\langle \Psi |H |\Psi\rangle}{\langle\Psi|\Psi\rangle}
\eeq
and can be shown to assume the following form\footnote{The derivation of Eq.(\ref{E_vmc}) is as follows
\beq
 E &=& \frac{\langle \Psi |H |\Psi\rangle}{\langle\Psi|\Psi\rangle}
= \frac{\sum _{|x\rangle}\langle \Psi|x\rangle \langle x | H |\Psi\rangle}{\sum_{|x\rangle}\langle\Psi|x\rangle \langle x|\Psi\rangle} 
= \sum_{|x\rangle}\frac{\langle \Psi|x\rangle}{\sum_{|x\rangle} |\langle x|\Psi\rangle|^2}  \sum_{|x'\rangle}  \langle x|H| x'\rangle \langle x'|\Psi\rangle \non
&=& \sum_{|x\rangle} \frac{|\langle \Psi|x\rangle|^2}{\sum_{|x\rangle}|\langle \Psi|x\rangle |^2} \sum_{|x'\rangle} \langle x|H|x'\rangle \frac{\langle x'|\Psi\rangle}{\langle x|\Psi\rangle}
=\sum_{|x\rangle} P_\text{loc}(x) E_\text{loc}(x) 
\eeq}
\beq
E =\sum_{|x\rangle} P_\text{loc}(x) E_\text{loc}(x) \label{E_vmc}
\eeq
where $P_\text{loc}(x)$ is the probability 
\beq
P_\text{loc}(x) = \frac{|\langle \Psi|x\rangle|^2}{\sum_{|x\rangle}|\langle \Psi|x\rangle |^2}
\eeq
of obtaining the local energy $E_\text{loc}(x)$, which is
\beq
E_\text{loc}(x) = \sum_{x'} \langle x|H|x'\rangle \frac{\langle x'|\Psi\rangle}{\langle x|\Psi\rangle}. \label{Eloc2}
\eeq
$E_\text{loc}(x)$ is non-zero only for those non-zero Hamiltonian elements $\langle x|H|x'\rangle \neq 0$. 
Using the notation $\langle x|\Psi\rangle = \Psi(x)$, we have
\beq
P_\text{loc}(x) = \frac{|\Psi(x)|^2}{\sum_{|x\rangle}|\Psi(x) |^2}, \hspace{5mm} E_\text{loc}(x) = \sum_{x'} \langle x|H|x'\rangle \frac{\Psi(x')}{\Psi(x)}. \label{Eloc}
\eeq

In the variational problem of interest, given a Hamiltonian $H$ and a trial wavefunction $\Psi$, we would like to estimate the ground state energy $E = \langle \Psi|H|\Psi\rangle$ in Eq.(\ref{E_vmc}) based on the local energy $E_\text{loc}$ given by Eq.(\ref{Eloc2}).
When the trial wavefunction  is a neural network quantum state $|\Psi(\vec\theta)\rangle$ with trainable parameters $\vec \theta$, at each training iteration $i$, the variational energy $E(\vec\theta)$ is calculated from the local energy $E_\text{loc}(\vec\theta)$ of the Monte Carlo samples generated from the NQS. The process of minimizing $E(\vec\theta)$ using an optimizer (such as SGD - Stochastic Gradient Descent or Adam - Adaptive Moment Estimation) updates the trainable parameters $\vec\theta$ until convergence is reached. 
\section{Hyperbolic RNN and GRU}\label{sec-hyprnn}
\subsection{Poincar\'e ball model of hyperbolic space } \label{sec-poincare}
In this section, we will summarize and recall the relevant formulas that were originally introduced in \cite{ganea-1805} to mathematically construct the hyperbolic RNN. The specific hyperbolic space model used in this construction is the Poincar\'e ball model $(\mathbb{D}^N, g^{\mathbb D})$ defined by the manifold $\mathbb{D}^N$
\beq
\mathbb{D}^N_c = \left\{ x\in \mathbb{R}^N: c||x||<1\right\} \label{eq-gyro}
\eeq
where the parameter $c$ is the Poincar\'e ball's radius. When $c=0$, $\mathbb{D}^N_c = \mathbb{R}^N$, while when $c>0$, $\mathbb{D}^N_c$ is the open ball of radius $1/\sqrt{c}$. In all the computations used in this work, we set $c=1$. 
The space $\mathbb{D}^N_c$ in Eq.(\ref{eq-gyro}) is equipped with the metric
\beq
g^{\mb D}_x = \lambda^2_x g^E, \hspace{5mm} \lambda_x = \frac{2}{1-||x||^2} \label{eq:poincare-metric}
\eeq
where $g^E = \mathbf{1}_N$ is the identity matrix representing the Euclidean metric. 
On a curved space, all arithmetic operations involving scalars and matrices (such as addition and multiplication) have to be redefined to take into account the geometry of the space. We will not repeat the derivation here (as the full details can be found in \cite{ganea-1805}) but simply quote the formulas that were actually used to define the hyperbolic RNN. In this section, only the most basic geometrical operations (such as addition, scalar multiplication, maps between hyperbolic space and its tangent space and parallel transport) will be defined since they are used to construct other operations such as pointwise nonlinearity and matrix multiplication, which will be defined in later sections in the order that they appear. The derivations and proofs for all formulas can be found in \cite{ganea-1805}.
\bei
\item The most basic operation in terms of which all other operations are defined is the Mobius addition $\oplus_c$ between scalars $x,y \in \mb{D}^N_c$. 
\beq
x\oplus_c y \equiv \frac{(1+2c\langle x, y\rangle + c||y||^2)x + (1-c||x||^2)y}{1 + 2c \langle x, y\rangle + c^2 ||x||^2 ||y||^2}\,. \label{eq-oplus}
\eeq
This operation is neither commutative nor associative but satisfies the following  $(-x) \oplus_c x = x\oplus_c(-x) = \mathbf{0}$. When $c=0$, the usual Euclidean addition operation is recovered $\lim_{c\ra 0} x\oplus_c y = x + y$. 
\item
For any point $x\in \mb{D}^N_c$, any vector $v\neq \mf{0}$ and any point $y\neq x$, the exponential and logarithmic maps $\exp^c_x: T_x\mb{D}^M_c \ra \mb{D}^M_c $ and $\log^c_x: \mb{D}^N_c \ra T_x\mb{D}^N_c$ between the hyperbolic space and its Euclidean tangent space are defined as
\beq
\exp_x^c(v) &=& x \oplus_c\lf[ \tanh\lf(\sqrt{c}\frac{\lambda_x^c||v||}{2}\rr)\frac{v}{\sqrt{c}||v||}\rr] \label{eq-exp}
\\
\log_x^c(y) &=& \frac{2}{\sqrt{c}\l_x^c} \tanh^{-1}\lf(\sqrt{c}||-x\oplus_c y||\rr) \frac{-x\oplus_cy}{||-x\oplus_c y||} \label{eq-log}
\eeq
where $\l_x^c = 2/(1-c||x||^2)$. 
Note that $\log_x^c$ and $\exp_x^c$ as defined in the above equations satisfies $\log_x^c(\exp_x^c(v)) = v$. When $x = \mf{0}$, the above maps take  more compact forms
\beq
\exp_\mf{0}^c(v) &=& \tanh\lf(\sqrt{c}||v||\rr) \frac{v}{\sqrt{c}||v||}\,\,\qquad 
\lf(T_{\mf{0}_M}\mb{D}^M_c \ra \mb{D}^M_c \rr) \label{eq-exp0}
\\
\log_\mf{0}^c(y) &=& \tanh^{-1}\lf(\sqrt{c}||y||\rr) \frac{y}{\sqrt{c}||y||}\qquad \lf(\mb{D}^N_c \ra T_{\mf{0}_N}\mb{D}^N_c\rr)\,. \label{eq-log0}
\eeq
Note that when $c=0$, we recover the Euclidean geometry limit $\lim_{c\ra 0} \exp_x^c(v) = x+v$ and $\lim_{c\ra 0}\log_x^c(y) = y-x$. 
\item 
The parallel transport $P^c_{\mf{0}\ra x}(v)$ of a vector $v\in T_\mf{0}\mb{D}^N_c$ to another tangent space $T_x\mb{D}^N_c$ can be defined in terms of the exponential and logarithmic maps of Eq.(\ref{eq-exp0}), Eq.(\ref{eq-log0}) as
\beq
P^c_{\mf{0}\ra x}(v) = \log_x^c\lf[x\oplus_c \exp_\mf{0}^c(v)\rr] \label{eq-pt}
\eeq
In terms of parallel transport, the Mobius addition for $x\in \mb{D}^N_c$ with $b\in \mb{D}^N_c$ can be written as
\beq
x\oplus_c b = \exp_x^c\lf[P_{\mf{0} \ra x}^c\lf(\log_\mf{0}^c(b)\frac{}{}\rr)\rr]\label{eq-m-pt}
\eeq
\item
The Mobius scalar multiplication of $x\in \mathbb{D}^N_c\backslash \{\mathbf{0}\}$ by $r\in \mathbb R$, denoted by $\otimes_c$, is defined as
\beq
r\otimes_c x = \frac{1}{\sqrt{c}} \tanh\lf(r \tanh^{-1}(\sqrt{c}||x||)\rr)\frac{x}{||x||}
\eeq
and $r\otimes_c \mathbf{0} = \mathbf{0}$. When $c=0$, the Euclidean scalar multiplication is recovered $\lim_{c\ra 0} r\otimes_c x = rx$. Interestingly, $r\otimes_c x$ can also be expressed in terms of the $\exp_\mf{0}^c$ and $\log_\mf{0}^c$ maps defined in Eqs.(\ref{eq-exp0}, \ref{eq-log0}) as
\beq
r\otimes_c x = \exp_\mf{0}^c\lf(r \,\log_\mf{0}^c(x)\rr), \qquad \forall r \in \mb{R}, x\in \mb{D}_c^N\,.
\eeq
\eni
\subsection{From Euclidean RNN to Hyperbolic RNN} \label{sec-hrnn}
Given an input vector $\vec x_i \in \mb{R}^{d_x}$ at step $i$ of size $d_x$, the defining equation for the Euclidean or conventional version of RNN is one that relates the hidden state vector $h_i\in \mb{R}^{d_h}$ of size $d_h$ to the input $x_i$ and the same hidden state vector at the previous time step $(i-1)$
\beq
\text{RNN} : && \vec h_i = f(W_h \vec h_{i-1} + U_h\vec x_{i} + \vec b_h)\,, \label{eq-rnn}
\eeq
where $W_h$ is a $d_h\times d_h$ weight matrix, $U_h$ is a $d_h\times d_x$ weight matrix, and $b_h\in \mb{R}^{d_h}$ is a vector of size $d_h$ known as the bias. Using Eq.(\ref{eq-rnn}), we can convert the Euclidean RNN to the hyperbolic RNN using
\beq
\text{Hyperbolic RNN} : && \vec h_i = f^{\otimes_c}\lf[(W_h \otimes_c \vec h_{i-1}) \oplus_c (U_h\otimes_c \vec x_{i}) \oplus_c \frac{}{}\vec b_h\rr] \label{eq-hrnn}
\eeq
where $W_h, U_i$ have the same meanings as in the Euclidean RNN case, but the hidden state vector $h_i \in \mb{D}^{d_h}_c$, input vector $\vec x_i \in \mb{D}^{d_i}_c$, and bias vector $b_h \in\mb{D}^{d_h}_c$ are all parts of the hyperbolic space. The operation $\oplus_c$ is defined in Eq.(\ref{eq-oplus}), while the matrix multiplication $\otimes_c$ and pointwise nolinearity $f^{\otimes_c}$ operations are defined in hyperbolic space as follows.
\ben

\item $f^{\otimes_c}: \mb{D}^N_c \ra \mb{D}^M_c$ is the Mobius-version of the pointwise Euclidean nonlinear activation function $f: \mathbb{R}^N \ra \mathbb{R}^M$
\beq
f^{\otimes_c}(x) \equiv \exp_\mf{0}^c\lf[f\lf\{\log_\mf{0}^c (x)\rr\}\rr]\,.
\label{eq-fc}
\eeq
where $\exp_\mf{0}^c$ and $\log_\mf{0}^c$ are defined in Eq.(\ref{eq-exp0}) and Eq.(\ref{eq-log0}). 
When $c=0$, we recover the Euclidean limit $\lim_{c\ra 0} f^{\otimes_c}(x) = f(x)$. 
\item The matrix-vector multiplication $\otimes_c$ acting on  the vector $x\in \mb{D}^N$ is the Mobius-version of the Euclidean matrix multiplication involving $W: \mb{R}^M \ra \mb{R}^N$ that acts on $x\in \mb{R}^N$ with $W x \neq \mf{0}$, and is defined as
\beq
W \otimes_c(x) = \frac{1}{\sqrt{c}}\tanh\left(\frac{||W x||}{||x||} \tanh^{-1}\lf(\sqrt{c}||x||\rr)\rr) \frac{W x}{||W x||}\,.\label{eq-otimes}
\eeq
Note that $W\otimes_c x = \mf{0}$ if $Wx =\mf{0}$.
\enn
Finally, we note that when the input $\vec x_i$ in Eq.(\ref{eq-hrnn}) is in Euclidean space,  we can still use the Eq.(\ref{eq-hrnn}) by making use of $\vec z_i = \exp_\mf{0}^c(\vec x_i)$\footnote{This is because using Eq.(\ref{eq-m-pt}) we have
\beq
\exp_{W\otimes_c \vec h_i}\lf[ P_{\mf{0} \ra W\otimes_c \vec h_i}^c (U\,x_i)\rr] = (W\otimes_c \vec h_i) \oplus_c \exp_0^c (U\,\vec x_i) = (W\otimes_c \vec h_i) \oplus_c (U\otimes_c\vec z_i)
\eeq} so that Eq.(\ref{eq-rnn})  becomes
\beq\vec h_i = f^{\otimes_c}\lf[(W_h \otimes_c \vec h_{i-1}) \oplus_c \frac{}{}(U_h\otimes_c \vec z_{i}) \oplus_c \vec b_h\rr] \,.\label{eq-hrnn-e}
\eeq
In this work, we will not make use of the RNN variant of hyperbolic RNNs, but its construction serves as an intermediate step in understanding the construction of the hyperbolic GRU variant, which is discussed in the next section. 
\subsection{From Euclidean GRU to Hyperbolic GRU} \label{sec-hgru}
GRU is a more sophisticated version of RNN with additional structures comprising the reset gate $r_i$ and the update gate $z_i$. The defining equations of the conventional or Euclidean GRU are
\beq
\text{GRU}: &&   \vec r_i = \s\lf(W_r   \vec h_{i-1} + U_r \vec x_i +   \vec b_r\rr) \label{gru-r}\\
            &&   \vec z_i = \s\lf(W_z   \vec h_{i-1} + U_z \vec x_i +   \vec b_z\rr) \label{gru-z} \\
            &&   \vec{\tilde h}_i = f\lf[W_h(  \vec r_i \odot  \vec h_{i-1}) + U_h\vec x_i +   \vec b_h\rr] \label{gru-ht}\\
 &&   \vec h_i = (1-z_i)\odot   \vec h_{i-1} +   \vec z_i\odot   \vec{\tilde h}_i \label{gru-update}
\eeq
In the above equations, $f$ is a nonlinear activation function, $\s$ is the sigmoid activation, $\vec x_{i}$ is the input vector of length $d_x$, $W_h, W_z, W_r$ are the $d_h\times d_h$ weight matrices, $U_h, U_z, U_r$ are the $d_h\times d_x$ weight matrices, $\vec b_h,   \vec b_z,   \vec b_r \in \mathbb{R}^{d_h}$ are the bias vectors, and $\odot$ is the pointwise multiplication operation.
\\\\
 The equations Eqs.(\ref{gru-r})-(\ref{gru-update}) above can be converted to the defining equation of the hyperbolic GRU using the hyperbolic operations $\oplus_c$, $\otimes_c$, $f^{\otimes_c}$ defined in Eqs.(\ref{eq-oplus}), (\ref{eq-otimes}), (\ref{eq-fc}) as follows.
 \bei
 \item First, the equations Eq.(\ref{gru-r}), Eq.(\ref{gru-z}) for the reset gate $\vec r_i$ and update gate $\vec z_i$ are updated to
 \beq
 \vec r_i &=& \s\,\log_\mf{0}^c\lf[(W_r\otimes_c  \vec h_{i-1}) \oplus_c (U_r\otimes_c \vec x_i) \oplus_c \frac{}{}  \vec b_r\rr] \non
  \vec z_i &=& \s\,\log_\mf{0}^c\lf[(W_z\otimes_c  \vec h_{i-1}) \oplus_c (U_z\otimes_c \vec x_i) \oplus_c \frac{}{}   \vec b_z\rr] \nonumber
 \eeq
 \item The term $\vec r_i\odot\vec h_{i-1}$ containing the pointwise multiplication operation $\odot$ is adapted to
 \beq
   \vec r_i\odot   \vec h_{i-1} \,\,\ra \,\,\text{diag}(r_i) \otimes_c   \vec h_{i-1}\,. \nonumber
 \eeq
 where $\text{diag}(r_i)$ denotes the diagonal matrix of size $d_h\times d_h$ with the diagonal entries being the entries of the vector $r_i$ (of length $d_h$). 
 \item The intermediate hidden state $\tilde h_i$ in Eq.(\ref{gru-ht}) is converted to
 \beq
 \vec{\tilde h}_i = f^{\otimes_c}\lf[(W\,\text{diag}(r_i)) \otimes_c \vec h_{i-1} \oplus (U\otimes_c \vec x_i)\frac{}{} \oplus_c  \vec b_h\rr] \nonumber
 \eeq
 \item The final GRU update equation, Eq.(\ref{gru-update}), is converted to
 \beq
\vec h_i = \vec h_{i-1} \oplus_c \,\text{diag}(z_i)\, \otimes_c\lf(-\vec h_{i-1} \oplus \vec{\tilde h}_i\rr)
 \eeq 
 \eni 
 Putting together all the update equations, we have the following defining equations for the hyperbolic GRU
 \beq
  \vec r_i &=& \s\,\log_\mf{0}^c\lf[(W_r\otimes_c \vec h_{i-1}) \oplus_c (U_r\otimes_c \vec x_i) \oplus_c \frac{}{}   \vec b_r\rr]\,, \label{hgru-r}
 \\
  \vec z_i &=& \s\,\log_\mf{0}^c\lf[(W_z\otimes_c \vec h_{i-1}) \oplus_c (U_z\otimes_c \vec x_i) \oplus_c \frac{}{}  \vec b_z\rr]\,, \label{hgru-z}
  \\
  \vec{\tilde h}_i &=& f^{\otimes_c}\lf[(W\,\text{diag}(r_i)) \otimes_c \vec h_{i-1} \oplus (U\otimes_c \vec x_i)\frac{}{} \oplus_c   \vec b_h\rr]\,, \label{hgru-ht}
  \\
  \vec h_i &=&\vec h_{i-1} \oplus_c \,\text{diag}(z_i)\, \otimes_c\lf(-\vec h_{i-1} \oplus \vec{\tilde h}_i\rr)\,.  \label{eq-hgru}
 \eeq
\subsection{Optimizing hyperbolic parameters} \label{sec-hyp-opt}
Hyperbolic RNN/GRU can be used in conjunction with  Euclidean Dense layers in a `plug-and-play' manner as noted by the authors of \cite{ganea-1805}. However, in going from hyperbolic to Euclidean layers, one must apply the exponential and logarithmic maps in Eq.(\ref{eq-exp0}) and Eq.(\ref{eq-log0}) accordingly. Furthermore, hyperbolic parameters must be optimized separately from their Euclidean counterparts using `hyperbolic' optimizer, which is chosen in this work to be RSGD - Riemannian Stochastic Gradient Descent \cite{rsgd}, \cite{rsgd-ganea}, \cite{rsgd-ganea-2}. To define RSGD, one starts from SGD, which is an update rule for a variable $x$ at time step $t$
\beq
x_{t+1} \leftarrow x_t - \alpha \,g_t \label{eq-sgd}
\eeq
where $g_t$ is the gradient of the objective/cost function $f_t$ and $\a>0$ is the learning rate. In a Riemannian manifold $\mc M$ equipped with metric $g$, $(\mc M, g)$, for smooth $f:\mc M\ra \mb{R}$, the Riemannian SGD is defined by the following update \cite{rsgd}
\beq
x_{t+1} \leftarrow \exp_{x_t}(-\a \,g_t) \label{eq-rsgd}
\eeq
where $\exp_{x_t}: T_{x_t}\mc M \ra \mc M$ is the exponential map defined in Eq.(\ref{eq-exp}), and $g_t \in T_{x_t}\mc M$ denotes the Riemannian gradient of $f_t$ at $x_t$. When $(\mc M, g) = (\mb{R}^n, \mf{I}_n)$, i.e. the $n$-dimensional Euclidean space, RSGD in Eq.(\ref{eq-rsgd}) reduces to SGD in Eq.(\ref{eq-sgd}) since $\exp_x(v) = x+v$ in Euclidean space.
\\\\
As a final remark in this section, we note that due to the many more mathematical operations involved in defining hyperbolic RNNs compared to their Euclidean counterparts, a drawback of these networks is that they are much more computationally intensive and take much longer to train than Euclidean RNNs. In this work, we only use the GRU variant of hyperbolic RNNs, and not the RNN variant (which tends to underperform GRU)\footnote{Throughout this work, however, sometimes we might use the term hyperbolic RNN interchangeably with hyperbolic GRU. This is the case, for example, in the term `RNN-based' where the RNN here can refer to either RNN/GRU (for Euclidean networks) or just GRU (for hyperbolic networks).}.  
\section{Euclidean/Hyperbolic RNN-based Neural Quantum State} \label{sec-nqs-rnn}
In this section, we describe in detail the RNN-based neural quantum state (NQS). In particular, the architecture of this type of NQS consists of a RNN/GRU unit, which can be either Euclidean or hyperbolic, and one or two dense layer(s),  depending on whether the NQS wavefunction is real or complex. The way the RNN-based NQS is defined and constructed here is almost the same as done in \cite{rnn_20}, with the important exception being the use of hyperbolic GRU. For Euclidean RNN-based NQS (both the GRU and RNN variants), the construction is the same as that of \cite{rnn_20} and Euclidean RNN-based NQS is included in this work to provide a performance benchmark for hyperbolic GRU-based NQS. 
\subsection{Real NQS wavefunction} \label{sec-wf-r}
For a discrete $N$-dimensional Hamiltonian systems in which the basis states $|\vec\sigma\rangle = (\sigma_1, \ldots, \sigma_N)$ with each component $\s_n$ $(1\leq n\leq N)$ assuming the discrete values $0,1, \ldots, (d_v-1)$\footnote{For example, a 2-level system has $d_v$ = 2 with two possible values : 0, 1. A 3-level system has $d_v=3$ with three possible values 0, 1, 2, and so on},
the RNN-based neural quantum state is defined as 
\beq
|\Psi\rangle = \sum_{\vec\sigma}\sqrt{P(\vec\sigma})|\vec\sigma\rangle \label{rnn_qs}
\eeq
where $P(\mathbf{\sigma})$ is the output generated by the RNN-based neural network\footnote{Applying the RNN quantum state Eq.(\ref{rnn_qs}) to the variational energy Eq.(\ref{Eloc2}), we have
\beq
E = \sum_{|\vec\s\rangle} \left(\frac{P(\vec\s)}{\sum_{|\vec\s\rangle} P(\vec\s)}\right) E_\text{loc}(\vec\s) \label{rnn_E}
\eeq
where the local energy of a generated sample $\vec\s$ is 
\beq
E_\text{loc}(\vec\sigma) = \sum_{|\vec \sigma'\rangle} \langle \vec\sigma |H |\vec\s'\rangle \frac{\sqrt{P(\vec\s')}}{\sqrt{P(\vec\s)}} \,.\label{rnn_Eloc}
\eeq}. $P(\sigma)$ is the probability of a particular configuration $|\vec\sigma\rangle$, and it is the product of the conditional probabilities $P(\s_n|\s_1\ldots \s_{n-1})$. 
\beq
P(\vec\sigma) = P(\s_1)P(\s_2|\s_3)\ldots\,P(\s_N|\s_1, \s_2, \ldots, \s_{N-1})
\eeq
The RNN-based neural network representing the NQS in this case consists of a RNN/GRU layer (either Euclidean or hyperbolic), and a dense layer with $d_v-1$ units with a Softmax activation function. 
The process of calculating $P(\vec\sigma)$ using this RNN-based NQS  is depicted in Fig.\ref{rnn-wavefunc-gen} and described step-by-step below.
\begin{figure}[H]
\centering
\begin{tikzpicture}[node distance = 1.3cm, thick]%
        \node[circle, draw] (0) {$P_1$};
        \node[rectangle, draw] (0b) [below of =0, yshift = -0.2cm]{$\begin{array}{c} \text{Dense} \\ (\text{Softmax})\end{array}$};
        \node[rectangle, draw](0c)[below of =0b, yshift=-0.6cm]{$\begin{array}{c}\text{Euclidean/}\\ \text{Hyperbolic}\\ \text{RNN}\\\end{array}$};
        \node[](0e)[left of = 0c, xshift = -1cm]{};
        \node[circle, draw](0d)[below of =0c, yshift = -0.3cm]{$\s_0$};

        \draw[->] (0d) -- node [right]{} (0c);
        \draw[->] (0c) -- node [right, xshift = 0.1cm]{$h_1$} (0b);
        \draw[->] (0b) -- node [right]{} (0);
        \draw[->] (0e) -- node [right,above, midway]{$h_0$} (0c);

        \node[circle, draw] (1) [right of = 0, xshift=2.2cm]{$P_2$};
        \node[rectangle, draw] (1b) [below of =1, yshift = -0.2cm]{$\begin{array}{c} \text{Dense} \\ (\text{Softmax})\end{array}$};
        \node[rectangle, draw](1c)[below of =1b, yshift=-0.6cm]{$\begin{array}{c}\text{Euclidean/}\\ \text{Hyperbolic}\\ \text{RNN}\\\end{array}$};
        \node[circle, draw](1d)[below of =1c,yshift = -0.3cm]{$\s_1$};

         \draw[->] (1d) -- node [right]{} (1c);
        \draw[->] (1c) -- node [right]{$h_2$} (1b);
        \draw[->] (1b) -- node [right]{} (1);
        \draw[->] (0c)-- node [right,above, midway]{$h_1$} (1c);

        \node[circle, draw] (3) [right of = 1, xshift = 2.2cm]{$P_3$};
        \node[rectangle, draw] (3b) [below of =3, yshift = -0.2cm]{$\begin{array}{c} \text{Dense} \\ (\text{Softmax})\end{array}$};
        \node[rectangle, draw](3c)[below of =3b, yshift=-0.6cm]{$\begin{array}{c}\text{Euclidean/}\\ \text{Hyperbolic}\\ \text{RNN}\\\end{array}$};
        \node[circle, draw](3d)[below of =3c,yshift = -0.3cm]{$\s_2$};

        \draw[->] (3d) -- node [right]{} (3c);
        \draw[->] (3c) -- node [right]{$h_3$} (3b);
        \draw[->] (3b) -- node [right]{} (3);
        \draw[->] (1c)-- node [right, above, midway]{$h_2$} (3c);

        \node[circle, draw] (4)[above of = 1] {$\Psi(\vec\sigma)$};
        \draw[->] (1) -- node [right]{} (4);
        \draw[->] (0) -- node [right]{} (4);
        \draw[->] (3) -- node [right]{} (4);
    \end{tikzpicture}
    \caption{Schematic of the process of calculating the  RNN wavefunction $\Psi(\vec\s)= \sqrt{P(\vec\s)}|\vec\s\rangle$ from the probability $P(\vec\s)$ of the sample $\vec\s$. Here $P(\vec\s) = P(\s_1)P(\s_2|\s_1)\ldots P(\s_N|\s_{N-1})$. For a compact representation, $N=4$ in the schematic. The RNN in the diagram can be either Euclidean RNN/GRU or Hyperbolic GRU. Figure adapted from \cite{rnn_20}.} \label{rnn-wavefunc-gen}
    \end{figure}
\bei
\item
At time step $i$, given an input $\vec\s_{i-1}$ (of dimension $d_v$) and a previous hidden state $\vec h_{i-1}$ of dimension $d_h$, the Euclidean/hyperbolic RNN maps it to a new hidden state $h_i$
\beq
\vec h_i = \texttt{RNN}(\vec h_{i-1}, \vec\s_{i-1})
 \label{hn}
\eeq
where the function \texttt{RNN}(\,.\,,\,.\,) is computed using Eq.(\ref{eq-rnn}), Eqs.(\ref{gru-r})-(\ref{gru-update}) for Euclidean RNN/GRU and Eqs.(\ref{eq-hrnn}), Eqs.(\ref{hgru-r})-(\ref{eq-hgru}) for hyperbolic RNN/GRU. The input $\vec x_i$ in those equations are now replaced by $\vec \s_i$, a one-hot encoded vector of length $d_v$.  When $i=1$, the hidden and initial states $\vec h_0$ and $\vec\s_0$ are both set to zero.
\\\\
If a hyperbolic RNN is used, we need to apply the logarithmic map Eq.(\ref{eq-log0}) on the RNN hidden state $\vec h_i$ before feeding it into the next layer, which is a normal (Euclidean) dense layer. 

\item The state $\vec h_i$ serves as an argument for the dense layer with a \texttt{Softmax} activation function to obtain the output $y_i$
\beq
\vec y_i  = \text{Softmax}\left(U\, \vec h_i + \vec c\rr), \hspace{10mm} \begin{dcases} U\in \mathbb{R}^{d_v \times d_h}: \,\text{weight matrix} \\ \vec c\in \mathbb{R}^{d_h}:\,\text{bias vector} \end{dcases}  \label{yn}
\eeq
and the \texttt{Softmax} function\footnote{which converts a vector of length K into a probability distribution of $K$ possible outcomes} is 
\beq
\text{Softmax}(v_k) = \frac{\exp(v_k)}{\sum_i \exp(v_i)} 
\eeq
$\vec y_i$ is a vector of length $d_v$ whose components sum up to 1. The probability $P(\s_i|\s_1\ldots \s_{i-1})$ is given by
\beq
P(\s_i|\s_1\ldots \s_{i-1}) = \vec y_i \,.\,\vec\s_i. 
\eeq
Note that $\vec\s_i$ is the one-hot encoded vector of length $d_v$, not to be confused with $\vec\sigma = (\s_1, \ldots, \s_{N})$, which is a vector of length $N$.

\item The total probability $P(\vec\sigma)$ appearing in Eq.(\ref{hn}) is then
\beq
P(\vec \sigma) = \prod_{n=1}^N \vec y_i\,.\,\vec\s_i\,.\label{Psigma}
\eeq
\eni
The process of generating the samples $\vec\s$ is similar to the process of calculating $P(\vec\s)$ above up to Eq.(\ref{yn}) when the output $\vec y_i$ at time step $i$ is obtained, after which $\s_i$ is sampled from $\vec y_i$, and then one-hot encoded and used as input, together with the hidden state $\vec h_{i-1}$, to repeat the process. 
Schematically, the generation of samples is shown in Fig.\ref{sample-gen}.
\begin{figure}[H]
\centering
\begin{tikzpicture}[node distance = 1.3cm, thick]%
        \node[circle, draw] (0) {$\s_1$};
        \node[circle, draw] (0a) [below of =0]{$y_1$};
        \node[rectangle, draw] (0b) [below of =0a, yshift = -0.2cm]{$\begin{array}{c} \text{Dense} \\ (\text{Softmax})\end{array}$};
        \node[rectangle, draw](0c)[below of =0b, yshift=-0.6cm]{$\begin{array}{c}\text{Euclidean/}\\ \text{Hyperbolic}\\ \text{RNN}\\\end{array}$};
        \node[](0e)[left of = 0c, xshift = -1cm]{};
        \node[circle, draw](0d)[below of =0c, yshift = -0.5cm]{$\s_0$};

        \draw[->] (0d) -- node [right]{} (0c);
        \draw[->] (0c) -- node [right]{$h_1$} (0b);
        \draw[->] (0b) -- node [right]{} (0a);
        \draw[->] (0a) -- node [right]{} (0);   
        \draw[->] (0e) -- node [right,above, midway]{$h_0$} (0c);

        \node[circle, draw] (1) [right of = 0, xshift=2.4cm]{$\s_2$};
        \node[circle, draw] (1a) [below of =1]{$y_2$};
        \node[rectangle, draw] (1b) [below of =1a,  yshift = -0.2cm]{$\begin{array}{c} \text{Dense} \\ (\text{Softmax})\end{array}$};
        \node[rectangle, draw](1c)[below of =1b, yshift=-0.6cm]{$\begin{array}{c}\text{Euclidean/}\\ \text{Hyperbolic}\\ \text{RNN}\end{array}$};
        \node[circle, draw](1d)[below of =1c,yshift = -0.5cm]{$\s_1$};

         \draw[->] (1d) -- node [right]{} (1c);
        \draw[->] (1c) -- node [right]{$h_2$} (1b);
        \draw[->] (1b) -- node [right]{} (1a);
        \draw[->] (1a) -- node [right]{} (1);
        \draw[->] (0c)-- node [right,above, midway]{$h_1$} (1c);
        \draw[rectangle connector= 1.5cm] (0)to (1d);

        \node[] (2)[right of = 1,xshift=1.3cm] {$\ldots$};
        \node[] (2c)[right of = 1c,xshift=1.3cm] {$\ldots$};
        \node[] (2d)[right of = 1d,xshift=1.3cm] {$\ldots$};
        \draw[->] (1c)-- node [right, above, midway]{$h_2$} (2c);
        \draw[rectangle connector= 1.4cm] (1)to (2d);

        \node[circle, draw] (3) [right of = 2, xshift=2.4cm]{$\s_N$};
        \node[circle, draw] (3a) [below of =3]{$y_N$};
        \node[rectangle, draw] (3b) [below of =3a, yshift = -0.2cm]{$\begin{array}{c} \text{Dense} \\ (\text{Softmax})\end{array}$};
        \node[rectangle, draw](3c)[below of =3b, yshift=-0.6cm]{$\begin{array}{c}\text{Euclidean/}\\ \text{Hyperbolic}\\ \text{RNN}\end{array}$};
        \node[circle, draw](3d)[below of =3c,yshift = -0.5cm]{$\s_{N-1}$};

        \draw[->] (3d) -- node [right]{} (3c);
        \draw[->] (3c) -- node [right]{$h_N$} (3b);
        \draw[->] (3b) -- node [right]{} (3a);
        \draw[->] (3a) -- node [right]{} (3);  
        \draw[->] (2c)-- node [right, above, midway]{$h_{N-1}$} (3c);
        \draw[rectangle connector= 1.9cm] (2)to (3d);

        \node[circle, draw] (4)[above of = 2, xshift = -1cm] {$\vec\sigma$};
        \draw[->] (1) -- node [right]{} (4);
        \draw[->] (0) -- node [right]{} (4);
        \draw[->] (3) -- node [right]{} (4);   
    \end{tikzpicture}
    \caption{The process of generating the length-$N$ samples $\vec\sigma = \lf(\s_1, \ldots, \s_{N}\rr)$ using an RNN-based neural network. Each entry $\s_i$ of $\vec\s$ is sampled from the output $y_i$ (generated by the RNN-Dense network using as input the previous entry $\s_{i-1}$ and the previous RNN hidden state $h_{i-1}$) and then one-hot encoded to be used as input, together with the previous hidden state $h_{i-1}$, for the generation of the next entry $\s_{i+1}$. For $i=0$, the input $\vec\s_0$ and first hidden state $h_0$ are initialized to zero. Figure adapted from \cite{rnn_20}.} \label{sample-gen}
    \end{figure}
\subsection{Complex NQS wavefunction} \label{sec-wf-c}
The complex RNN-based NQS wavefunction has a phase factor $\phi(\vec\s)$ in addition to a real amplitude $P(\vec\s)$, and is defined as
\beq
\Psi(\vec\s) = \sum_{\vec\s} \exp(i\phi(\vec\s))\,\sqrt{P(\vec\s)}\,|\vec\s\rangle\, \label{eq-wf-c}
\eeq
The architecture of the RNN-based neural network representing the complex NQS differs from that of the RNN-based NQS by an additional Dense layer, with the Softsign function, modeling the phase. The schematic of the complex RNN-based NQS is shown in Fig.\ref{rnn-wavefunc-gen-c}. 
In Eq.(\ref{eq-wf-c}), the amplitude $P(\vec\s)$ is generated  using a Softmax layer in the same way as in the case of the real NQS wavefunction, i.e.
\beq
P(\vec\s) = \prod_{i=1}^N P(\vec\s_i), \qquad P(\vec\s_i)  = y^{(1)}_i \,.\,\vec\s_i, \qquad y^{(1)}_i = \text{Softmax}\lf(U\,h_i + c\rr) \label{eq-p}
\eeq
while the phase $\phi(\vec\s$) is generated using a Softsign layer. The Softsign function is defined as
\beq
\text{Softsign}(x) = \frac{x}{1+|x|} \in (-1,1)\,.
\eeq
The total phase factor  $\phi(\vec\s)$ of Eq.(\ref{eq-wf-c})  is the sum of the phase factors $\phi_i$ resulting from each entry $\s_i$ of $\vec\s$
\beq
\phi(\vec\s) = \sum_{i=1}^N\phi_i, \qquad \phi_i = y_i^{(2)}\,.\,\vec\s_i\,, \label{eq-phi}
\eeq
where the output $\vec y^{(2)}_i$ is obtained from the Softsign layer
\beq
\vec y^{(2)}_i= \pi\, \text{Softsign}\lf(U^{(2)}\vec h_i + c^{(2)}\rr)
\eeq
In Eqs.(\ref{eq-p}), (\ref{eq-phi}), we note that $\vec\s_i$ is the one-hot encoded vector with dimension $d_v$, and it is the $i^{th}$ entry of the sample $\vec\s = (\s_1,\ldots, \s_N)$.
\begin{figure}[H]
\centering
\begin{tikzpicture}[node distance = 1.3cm, thick]%
        \node[circle, draw] (0) {$P_1$};
        \node[circle, draw] (0a)[right of = 0] {$\phi_1$};
        \node[circle, draw] (0b) [below of =0]{$S$};
        \node[circle, draw] (0b2) [below of =0a]{$SS$};
        \node[rectangle, draw](0c)[below of =0b, xshift = 0.5cm, yshift=-0.4cm]{$\begin{array}{c}\text{Euclidean/}\\ \text{Hyperbolic}\\ \text{RNN}\\\end{array}$};
        \node[](0e)[left of = 0c, xshift = -1cm]{};
        \node[circle, draw](0d)[below of =0c, yshift = -0.4cm]{$\s_0$};

        \draw[->] (0d) -- node [right]{} (0c);
        \draw[->] (0c) -- node [right]{$h_1$} (0b);
        \draw[->] (0c) -- node [right]{} (0b2);
        \draw[->] (0b) -- node [right]{} (0);
        \draw[->] (0b2) -- node [right]{} (0a);
        \draw[->] (0e) -- node [right,above, midway]{$h_0$} (0c);

        \node[circle, draw] (1) [right of = 0a, xshift=0.3cm]{$P_2$};
        \node[circle, draw] (1a) [right of = 1]{$\phi_2$};
        \node[circle, draw] (1b) [below of =1]{$S$};
        \node[circle, draw] (1b2) [below of =1a]{$SS$};
        \node[rectangle, draw](1c)[right of =0c, xshift=1.9cm]{$\begin{array}{c}\text{Euclidean/}\\ \text{Hyperbolic}\\ \text{RNN}\\\end{array}$};
        \node[circle, draw](1d)[below of =1c,yshift = -0.4cm]{$\s_1$};

        \draw[->] (1d) -- node [right]{} (1c);
        \draw[->] (1c) -- node [right]{$h_2$} (1b);
        \draw[->] (1b) -- node [right]{} (1);
        \draw[->] (1b2) -- node [right]{} (1a);
         \draw[->](1c) -- node [right]{} (1b2);
        \draw[->] (0c)-- node [right,above, midway]{$h_1$} (1c);

        \node[circle, draw] (3) [right of = 2,xshift = -0.8cm]{$P_3$};
        \node[circle, draw] (3a) [right of = 3]{$\phi_3$};
        \node[circle, draw] (3b) [below of =3]{$S$};
        \node[circle, draw] (3b2) [below of =3a]{$SS$};
        \node[rectangle, draw](3c)[right of = 1c, xshift=1.9cm]{$\begin{array}{c}\text{Euclidean/}\\ \text{Hyperbolic}\\ \text{RNN}\\\end{array}$};
        \node[circle, draw](3d)[below of =3c,yshift = -0.4cm]{$\s_2$};

        \draw[->] (3d) -- node [right]{} (3c);
        \draw[->] (3c) -- node [right]{$h_3$} (3b);
        \draw[->] (3b2) -- node [right]{} (3a);
        \draw[->] (3b) -- node [right]{} (3);
        \draw[->](3c) -- node [right]{} (3b2);
        \draw[->] (1c)-- node [right, above, midway]{$h_2$} (3c);

        \node[circle, draw] (4)[above of = 1, xshift = 1.0cm, yshift = 0.8cm] {$\Psi(\vec\sigma)$};
        \draw[->] (1) -- node [right]{} (4);
        \draw[->] (1a) -- node [right]{} (4);
        \draw[->] (0) -- node [right]{} (4);
        \draw[->] (0a) -- node [right]{} (4);
        \draw[->] (3) -- node [right]{} (4);
        \draw[->] (3a) -- node [right]{} (4);

    \end{tikzpicture}
    \caption{Schematic of the calculation of the RNN wavefunction  $\Psi(\vec\s) = \sum_{\vec\s}\exp(i\phi(\vec\s) \sqrt{P(\vec\s)}|\vec\s\rangle$ where the amplitude $P(\vec\s) = P(\s_1)P(\s_2|\s_1)\ldots P(s_N|\s_{N-1})$ and the phase $\phi(\vec\s) = \sum_{i=1}^N \phi(\vec\s_i)$. For ease of illustration, $N$ is chosen to be 3. (S) and (SS), correspondingly, denote the dense layer with the Softmax and Softsign activation function. Figure adapted from \cite{rnn_20}. } \label{rnn-wavefunc-gen-c}
    \end{figure}
In the experiments involving the complex NQS wavefunction in this work, the Marshall sign \cite{j1j2-marshall} is  used (since it leads to some improvements in the results, as reported by \cite{rnn_20}). The wavefunction with the Marshall sign, $\Psi_M(\vec\s)$, can be written as
\beq
\Psi_M(\vec\s) = (-1)^{M_A(\vec\s)} \Psi(\vec \s)\label{eq-marshall}
\eeq
where $M_A(\vec\s) = \sum_{i\in A} \s_i$, with $A$ comprising the sites belonging to the sublattice of all even or all odd sites in the lattice. The factor $(-1)^{M_A(\vec\s)}$ is the Marshall sign of the wavefunction, and $\Psi(\s)$ is the NQS wavefunction before the Marshall sign was applied (for complex NQS, this comprises both the amplitude and phase parts).  Eq.(\ref{eq-marshall}) is exact in unfrustrated spin systems when only neareast neighbor interactions $J_1$ (among the spin sites) are nonzero. When next-nearest (or higher degrees of) neighbor interactions ($J_2$ or $J_{i\geq 2}$) are nonzero, Eq.(\ref{eq-marshall}) is no longer exact but still holds approximately \cite{cnn-1903}.
\section{VMC with hyperbolic RNN-based NQS} \label{sec-res}
In this section, we report the results of running various VMC experiments for several quantum many-body systems using the GRU variant of hyperbolic RNNs as well as the Euclidean RNN/GRU as ansatzes for the trial ground state wavefunction. In particular, we choose the following four prototypical systems as settings for our VMC experiments: 
\bei
\item One-dimensional Transverse Field Ising model (1D TFIM) - Section \ref{sec-res-tfim}
\item Two-dimensional Transverse Field Ising model (2D TFIM) - Section \ref{sec-res-2dtfim}
\item One-dimensional Heisenberg $J_1J_2$ model (1D $J_1J_2$) - Section \ref{sec-res-j1j2}
\item One-dimensional Heisenberg $J_1J_2J_3$ model (1D $J_1J_2J_3$) - Section \ref{sec-res-j1j2j3}
\eni
These systems are the most basic testing grounds for different types of NQS ansatzes, and they have been used routinely in the literature. As a benchmark  against which to evaluate the performances of all NQS ansatzes, we will use DMRG (Density Matrix Renormalization Group). 
With 1D systems, DMRG results are considered to be exact, while in 2D systems, this is no longer the case, but this is still a reliable benchmark to use since DMRG provides state-of-the-art computational results for various condensed matter systems. 
\\\\
As the performances of Euclidean RNN-based NQS have been established in the work \cite{rnn_20} for the 1D TFIM, 2D TFIM and 1D $J_1J_2$  systems, we do not aim to replicate their results but instead to establish the performances of hyperbolic GRU-based NQS using the performances of Euclidean RNN/GRU-based NQS as benchmarks in settings of smaller scales, due to the fact that hyperbolic networks are more computationally intensive and take much longer to train than their Euclidean counterparts. Specifically, for the 1D TFIM , the work \cite{rnn_20} explored settings up to $N=1000$ spins, while we restrict the number of spins in our settings up to $N=100$. For the 2D TFIM, \cite{rnn_20} studied a square lattice of size $(N_x, N_y) = (12, 12)$ while we restrict the lattice size to $(N_x, N_y) = (9,9)$. For the $J_1J_2$ systems, \cite{rnn_20} used a setting involving $N=100$ spins while we restrict the number of spins in our setting to $N=50$. The 1D $J_1J_2J_3$ model was not studied in \cite{rnn_20}. 
\\\\
Before moving further, several important remarks are in order. These concern the different aspects of the experiments that remain unchanged throughout this work (unless otherwise explicitly stated). 
\bei
\item  During the training process of all neural networks in this work, we fixed  the number of generated spin configuration samples to be 50. At each iteration of the training process, a neural network ansatz generates 50 samples (spin configurations) whose mean energy is used to approximate the true ground state energy of the Hamiltonian system under study. Convergence is reached when there is no noticeable change in the mean energy and the energy variance of all the samples is smaller than a certain tolerance threshold\footnote{This specific values of the tolerance threshold vary for different VMC experiments, and can range from as low as 0.05 to 1.0 for the 1D TFIM case. In other settings such as 2D TFIM, 1D $J_1J_2$ and 1D $J_1J_2J_3$ models, the variance tolerance can be much higher.}. 
This choice of 50 samples (during training) was made based on the known fact that RNN-based, autoregressive NQS do not require too many samples to achieve convergence since an autoregressive architecture allows for an exact and independent sampling process by forward-generating spins from conditional distributions (with no Markov chain Monte Carlo sampling required) \cite{1902-autoregressive}, \cite{rnn_20}. Furthermore, we also observed during our experiments that increasing the number of samples generated to 100 or 200 did not really improve the results while exponentially increasing the training time\footnote{For Euclidean RNN/GRU, we had no problem using 200 samples during the training process, and 200 was in fact the number of samples used in the work \cite{rnn_20}, but for hyperbolic GRU, this took too long with the limited computing resources that we currently have. Already with 50 samples, the training of hyperbolic GRU took much longer than that of their Euclidean counterparts.}.  In the inference process, however, when the trained neural networks were loaded to generate new samples to estimate the energy, we used a significantly higher  number (in the order of thousands) of samples to estimate the mean energy. 
\item

For Euclidean ansatzes, we used \texttt{Adam} optimizer with an exponential decay learning rate while for the hyperbolic ansatz, we used RSGD (Riemannian SGD) with the fixed learning rate of $\a= 10^{-2}$ for hyperbolic parameters and the same \texttt{Adam} with exponentially decaying learning rate for Euclidean parameters. In the hyperbolic GRU ansatz, only the biases ($b_h, b_r, b_z)$ are hyperbolic, while the weight matrices are Euclidean. This hyperbolic network configuration is the same as the one used in \cite{ganea-1805}, and is kept unchanged throughout this work.
\item
In a typical training process, the weights of the neural network model under training are constantly overwritten after each training epoch, so the final model is the one saved at the last epoch in the training. Theoretically, at convergence when the model reaches a constant value without any further changes, the weights of the model stop changing and the weights at the last epoch are the final weights of the model. In practice, what is observed is that even at convergence when the model has reached the constant energy level, there are still minor fluctuations within a very small range of this energy, espcially with this kind of training involving the generation of samples. To address this issue, in our training process, the saving of the model weights at any particular training epoch is conditioned on the fact that the model has actually improved (the criteria for judging the improvement, programmed in the training loop, include not only a lower mean energy but also a lower variance, subject to a certain variance tolerance threshold) compared to the previous training epoch. As such, when convergence is reached in a practical sense (in the sense that the mean energy has reached a constant level albeit with fluctuations within a very narrow range), our last saved model correponds to the best possible result, i.e. a model that has already converged with the lowest energy and the variance smaller than the tolerance threshold. This means that our final model might not necessarily be the model at the last epoch in the training, but the model saved at some earlier epoch near the end of the training\footnote{We note that the practice of saving the best model only after a practical convergence is reached is commonly used in all types of neural network training not restricted to the present context of NQS. In Keras (\href{https://keras.io/api/callbacks/model_checkpoint/}{keras.io/api/callbacks/model\_checkpoint/}), one can choose to save the best model only (e.g. one that has the lowest validation loss or one that fulfills a certain set of custom defined criteria) as the training progresses using a function built-in in \texttt{ModelCheckpoint} callback by selecting the option \texttt{save\_best\_only = True}. }. In other words, the training loop might continue until the specified number of iterations is reached but unless a better result is obtained, the final model is the model saved at the epoch where the best result was obtained\footnote{Our training results are perfectly reproducible, as long as the same random seed parameter is chosen. }. For example, with the total number of epochs set at 450, if the convergence is reached after 350 epochs (the energy  no longer changes noticeably but varies within a very narrow range of order 0.01 to 0.1), and the best model (with the lowest energy and the variance below the tolerance threshold) is the one at epoch 375, training will still carry on until epoch 450 but no new model is saved after epoch 375 unless there is an improved model at the $i^{th}$ epoch where $375<i<450$.
\eni
\subsection{1D Transverse Field Ising Model}\label{sec-res-tfim}
The Hamiltonian of the one-dimensional transverse field Ising model (1D TFIM) with open boundary condition is
\beq
H_\text{1DTFIM} = -J\sum_{i=1}^{N-1} \s_i^z \s^z_{i+1} - B\sum_{i=1}^N \s^x_i
\eeq
where $\s_x, \s_y, \s_z$ are the Pauli matrices. The first sum in the Hamiltonian runs over pairs of nearest neighbors, while the second sum runs over all sites in the 1D spin chain. This system is known to exhibit a phase transition at the critical value of $B=1$ (with $J=1$) from a disordered (paramagnetic) to an ordered (ferromagnetic) state. 
\subsubsection{Experiment details}
For the 1D TFIM system, we choose four settings with the number of spins $N$ being 20, 40, 80 and 100. Three different RNN-based real NQS, namely Euclidean RNN, Euclidean GRU and hyperbolic GRU, are used as ansatzes in the VMC experiments for 1D TFIM at each $N$ (see Table \ref{tab:1dtfim_setting}). The architecture of these ansatzes is the same as that of the real NQS wavefunction described in Section \ref{sec-wf-r}, which consists of an RNN (either Euclidean or hyperbolic) with the hidden dimension $d_h=50$ and a Dense layer with 2 units (representing the 0 or 1 entry of the spin $\s_i$ at site $i$) with the Softmax activation function. The wavefunction is given by Eq.(\ref{rnn_qs}) in which the state $|\vec\s\rangle$ is a vector of length $N$ with each entry assuming a value of either 0 or 1 (spin up or down). 
\begin{table}[!h]
\centering
\begin{tabular}{|l|c|cc|}
\hline
Ansatz & Name  &Parameters& \\ \hline
Euclidean RNN & \texttt{eRNN-50-s50} & 2752 &\\
Euclidean GRU & \texttt{eGRU-50-s50} & 8052 &\\
Hyperbolic GRU & \texttt{hGRU-50-s50}  & 8052 &
\\\hline
\end{tabular}
\caption{The variants of NQS ansatzes used for the VMC experiments involving the 1D TFIM Hamiltonian with various system sizes $N=20,40,80, 100$ for $J=1$ and $B = 1$.} \label{tab:1dtfim_setting}
\end{table}
The number of parameters of the Euclidean RNN ansatz is 2752, while the number of parameters of the Euclidean and hyperbolic GRU-based ansatz is 8052. For all three types of ansatzes at all $N$, convergence is reached very quickly at around 100 epochs with the total number of training steps set at 120 epochs. Even with this very small number of epochs, for the hyperbolic GRU to complete the training loop, it takes more than 1.5 hours for $N=20$ and up to 20 hours for $N=100$ on an average laptop\footnote{An average laptop at the time of writing is one with around 8-16Gb of RAM} with no GPU. On the other hand, Euclidean RNN and GRU take around 0.1 hours for $N=20$ and around 1.5 hours for $N=100$ on the same hardware.
\subsubsection{Results}
In  Table \ref{tab:1dtfim_res}, we list the  inference results,  obtained by using the trained models to generate new spin configurations to calculate the new NQS wavefunction, for the  VMC experiments for the 1D TFIM at $N=20, 40, 80, 100$. The number of samples used for inference is $10^4$.  The convergence plots for the mean energy recorded from the training process are shown in Fig.\ref{fig_1dtfim_n20_40_80} in Appendix \ref{sec-app-1dtfim}. 

\begin{table}[!ht]
\centering
\begin{tabular}{lcc ccc}
\hline\hline
Ansatz &  $N=20$  & $N=40$ & $N=80$ & $N=100$ &
\\\hline\hline
\texttt{eRNN-50-s50} &  25.0742 & -50.5011 &   -101.3740  & -126.8042 &\\
& (0.0030)  &  (0.0050)  &(0.0050) & (0.0060) &
\\ \hline
\texttt{eGRU-50-s50} & -25.0838 & \textbf{-50.5415} &   \textbf{-101.4131}   &-126.8717 &\\
& (0.0030 ) &  (0.0030) &  (0.0040) &  (0.0040) &
\\\hline
\texttt{hGRU-50-s50} &  \textbf{-25.0874}  &  -50.5013  &  -101.3949  & \textbf{-126.8763} &\\
& (0.0020) &   (0.0050) &  (0.0060) &  (0.0050) &
\\
\hline
Exact (DMRG)&  -25.1078&  -50.5694&  -101.4974 & -126.9619 &\\
\hline
\end{tabular}
\caption{VMC inference results (using 10000 samples) for 1D TFIM Hamiltonian with various system sizes $N=20,40,80, 100$ for $J=1$ and $B = 1$. For each ansatz, the mean $E$ is listed first, together with the corresponding standard error directly below (in brackets). These results are obtained from the inference process in which the trained parameters are loaded into new models and these models are used to calculate the mean energy.  The best inference result for the converged mean energy at each $N$ is noted in bold.} 
\label{tab:1dtfim_res}
\end{table}
\begin{figure*}[ht!]
    \centering
    \includegraphics[width = .6\textwidth]{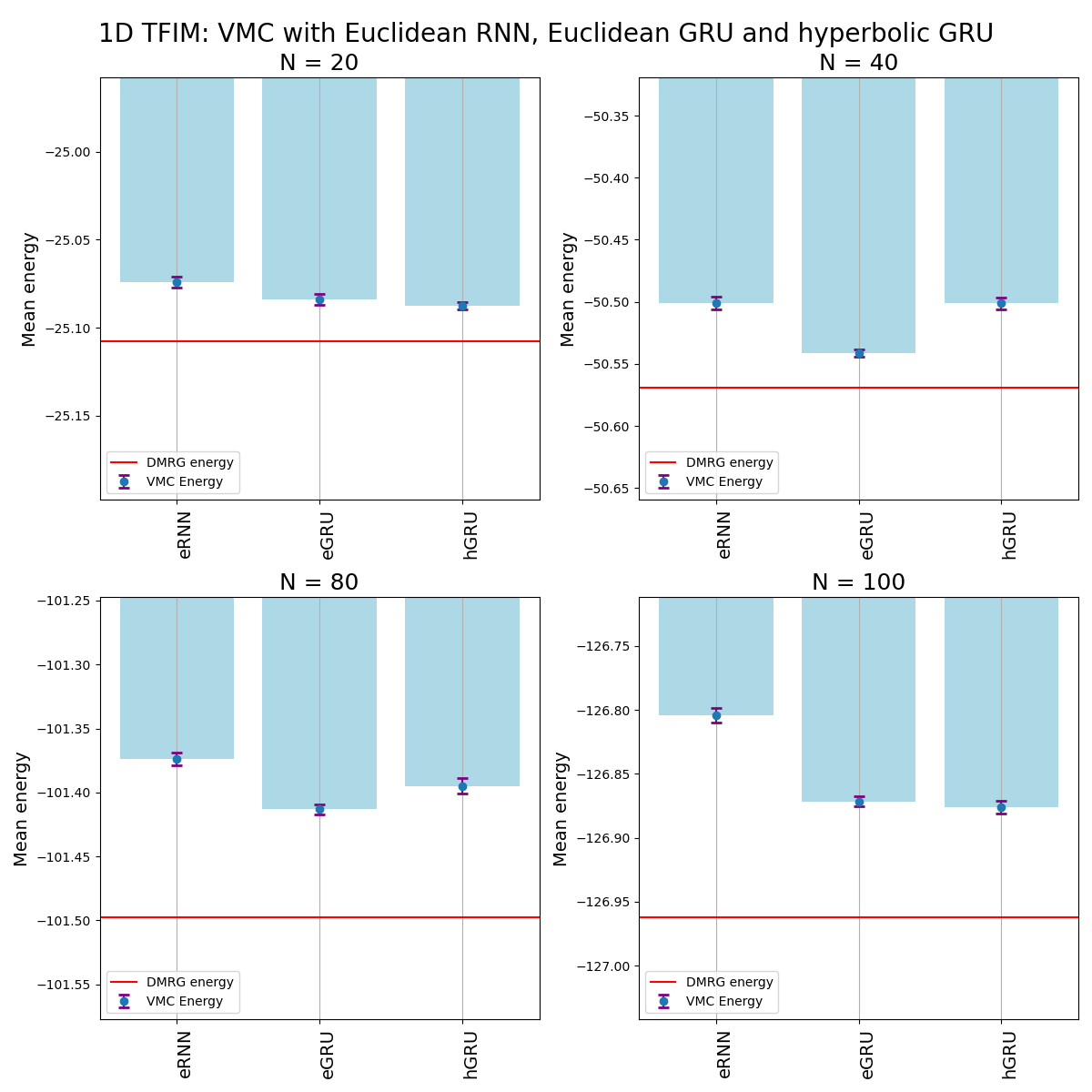}
\caption{VMC experiments for 1D TFIM with $N=20,40, 80$ and 100 spins (from top to bottom row, left to right): Comparisons of the performances of the variants \texttt{eRNN-50-s50}, \texttt{eGRU-50-s50}, \texttt{hypGRU-50-s50} (listed in Table \ref{tab:1dtfim_setting}) - abbreviated as \texttt{eRNN}, \texttt{eGRU} and \texttt{hGRU} on the horizontal axis of each subfigure. In each subfigure, the mean energy for each NQS ansatz is shown as a dot with an error bar (representing the standard error). }
\label{fig_1dtfim_comparison}
\end{figure*}
The results listed in  Table \ref{tab:1dtfim_res} are graphically presented in Fig.\ref{fig_1dtfim_comparison}, from which one can see immediately the performances of the three types of NQS ansatzes with respect to one another. In all cases, Euclidean GRU and hyperbolic GRU NQS outperformed Euclidean RNN NQS. However, when it comes to the comparing the performances of Euclidean and hyperbolic GRUs, the situation is not as clear-cut.
\bei
\item
For $N=20, 100$, the best NQS ansatz is \texttt{hGRU-50-s50}, with the second best one being \texttt{eGRU-50-s50}.
\item 
For $N=40, 80$, the best NQS ansatz is \texttt{eGRU-50-s50}, with the second best one being \texttt{hGRU-50-s50}.
 \eni

\FloatBarrier
\subsection{2D Transverse Field Ising Model}\label{sec-res-2dtfim}
In this section, we consider the two-dimensional transverse field Ising model on a square lattice with open boundary condition, with the following Hamiltonian 
\beq
H = -J\sum_{\langle i,j\rangle} \s_z^i \s_z^{j} - B \sum_i \s_x^i
\eeq
where the sum $\langle i,j\rangle$ is over pairs of nearest-neighbor on the square 2D lattice. This system has a phase transition at the critical value of $B_c = 3.044$ separating the magnetically ordered phase from a random paramagnet \cite{2d-tfim-pc}. In our VMC experiments, as is the case in \cite{rnn_20}, we fix $J=1.0$ and $B = 3.0$. 
\subsubsection{Experiment details}\label{sec-res-2dtfim-exp}
For the 2D TFIM, we will perform the VMC experiments using three different NQS ansatzes: one-dimensional Euclidean GRU, one-dimensional hyperbolic GRU and two-dimensional Euclidean RNN (see Appendix \ref{sec-2d-rnn} for the detailed description of the 2D RNN neural network and the sample generation process involving both 1D and 2D NQS), the details (including the names and number of parameters) of which are included in Table \ref{tab:2d_tfim_setting}. Since the 2D Euclidean RNN NQS was specially designed by taking into account the 2D geometry of the TFIM physical system, it is our expectation that 2D RNN will outperform both 1D Euclidean and hyperbolic GRU. The goal of this experiment, therefore, is not to determine the best, but rather the second-best performing NQS. In other words, \textit{what we are really interested in, and would like to find out is the performance of 1D hyperbolic GRU with respect to 1D Euclidean GRU.} The inclusion of the 2D RNN NQS was made to provide an overall benchmark in the 2D TFIM case.
\\\\
 The size $(N_x, N_y)$ of the square lattice in the 2D TFIM is chosen to be (5,5), (7,7), (8,8) and (9,9)\footnote{Unfortunately, we do not have enough computing resources to carry out the VMC experiments for the case of $(N_x, N_y) \geq (10,10).$} corresponding to systems with 25, 49, 64 and 81 spins. All neural network ansatzes are real NQS wavefunction (described in Section \ref{sec-wf-r}) whose architecture comprises an RNN unit (either Euclidean GRU, hyperbolic GRU or 2D Euclidean RNN) with the hidden dimension $d_h=50$, and a Dense layer with the Softmax activation function. 
\begin{table}[!ht]
\centering
\begin{tabular}{|l|c|c|}
\hline
Ansatz  & Name  &  Parameters\\
\hline
1D Euclidean GRU & \texttt{1d-eGRU-50-s50} & 8052\\
\hline
1D Hyperbolic GRU &\texttt{1d-hGRU-50-s50} &  8052\\
\hline 
2D Euclidean RNN & \texttt{2d-RNN-50-s50} & 5352
\\
\hline
\end{tabular}
\caption{The  variants of Euclidean and hyperbolic GRU used as NQS ansatzes to run the VMC experiments involving the 2D TFIM  with $(N_x, N_y) = (5,5), (7,7), (8,8), (9,9)$, at $J = 1$ and $B=3.0$. } \label{tab:2d_tfim_setting}
\end{table}

Other training details are largely the same as described in the 1D TFIM, and will not be repeated here. For $(N_x, N_y) = (5,5), (7,7)$, with the number of training epochs being 450, 1D Euclidean GRU takes less than 2 hours, while 1D hyperbolic GRU takes up to 14 hours to finish the training. For $(N_x, N_y)=(8,8), (9,9)$, with the number of training epochs reduced to 350 due to the significantly longer running time required, 1D Euclidean GRU took around 4 hours, while  1D hyperbolic GRU could take up to around 50 hours to finish the training. For the lighter 2D Euclidean RNN that reached convergence faster than the 1D ansatzes, the numbers of epochs varied and were 200 for $(N_x, N_y) = (5,5)$, 300 for $(N_x, N_y)=(7,7)$ and 350 for $(N_x, N_y)=(8,8), (9,9)$. The training times of 2D Euclidean RNN ranged from 0.2 hour for $(N_x, N_y) = (5,5)$ to 3.5 hours for $(N_x, N_y)= (9,9)$.
\subsubsection{Results}\label{sec-res-2dtfim-res}
The inference results (using 10000 samples) for the VMC experiments involving the 2D TFIM setting are listed in Table \ref{tab:2dtfim_res_int}. The convergence curve plots for the mean energy are shown in Fig.\ref{2d-tfim-meanE} in Appendix \ref{sec-app-2dtfim}.

\begin{figure}[!h]
\centering
\includegraphics[width = .6\textwidth]{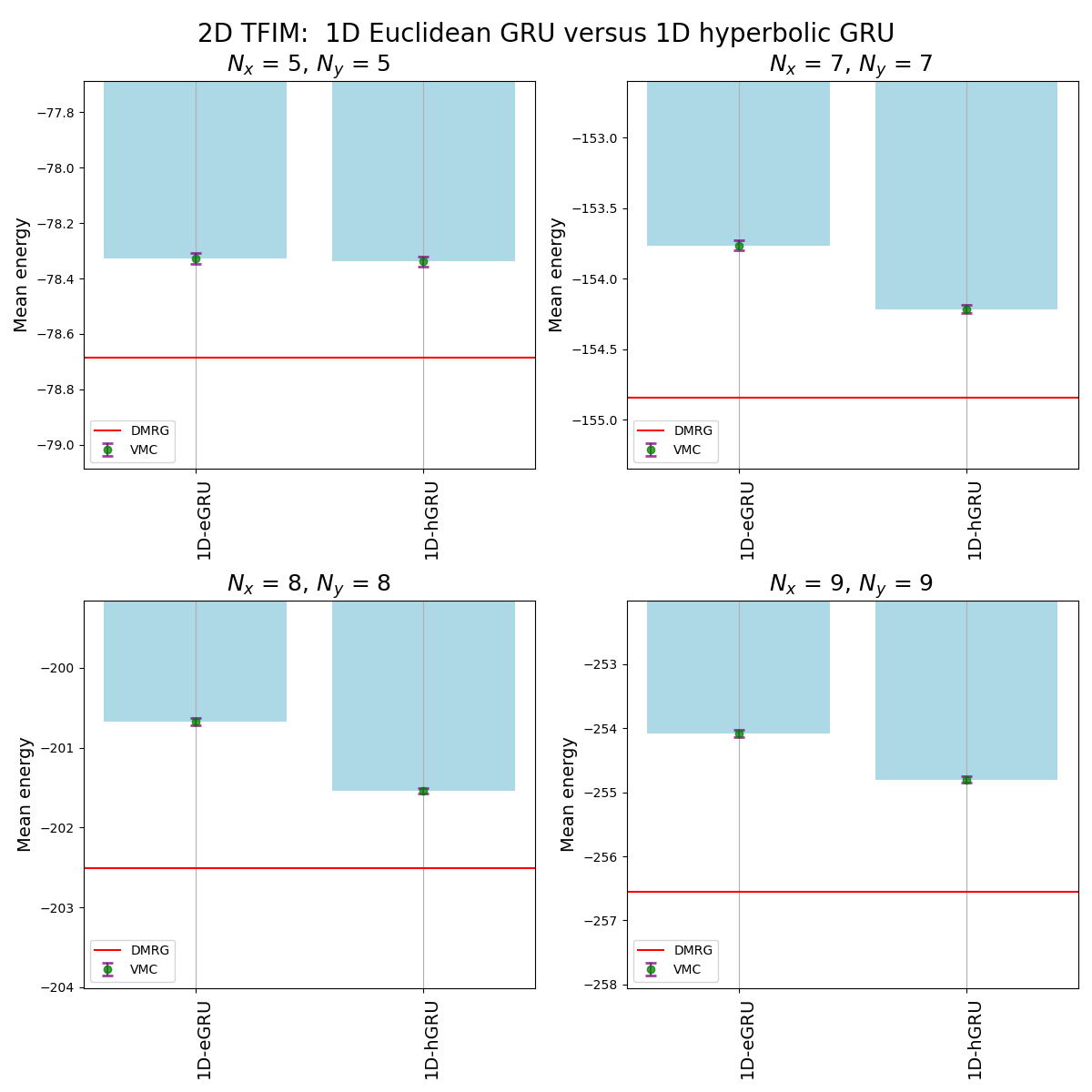}
\caption{Comparisons of the performances of 1D Euclidean GRU and 1D hyperbolic GRU ansatzes listed in Table \ref{tab:2d_tfim_setting} for the 2D TFIM with different ($N_x, N_y$) lattices - clockwise from top: $(N_x, N_y) = (5,5), (7,7), (9,9), (8,8)$. Each dot with error bar represents the mean $E$ value of an NQS ansatz. For all cases, 2D Euclidean RNN is the best ansatz and is not included in this comparison, which strictly measures the performances of the 1D Euclidean versus 1D hyperbolic GRU. The second best ansatz is always 1D hyperbolic GRU for all $(N_x, N_y)$.}\label{2d-tfim-comparison}
\end{figure}

\begin{table}[!ht]
\centering
\begin{tabular}{lcc ccc}
\hline\hline
Ansatz &  $(N_x, N_y)=(5,5)$  & $(N_x, N_y)=(7,7)$ & $(N_x, N_y)=(8,8)$ & $(N_x, N_y)=(9,9)$ 
\\\hline\hline
\texttt{1d-eGRU-50-s50}  &  -78.3274  &  -153.7647  & -200.6790  & -254.0806 & \\
& (0.0190) &  (0.0370) &  (0.0450) &  (0.0570) &\\
\hline
\texttt{1d-hGRU-50-s50} & \textbf{-78.3385} & \textbf{-154.2155}  & \textbf{-201.5419}  & \textbf{-254.8021} &  \\
& (0.0190) &   (0.0280) &  (0.0330) &  (0.0480)& \\
\hline
\texttt{2d-eRNN-50-s50}& \texttt{-78.6554} &   \texttt{-154.7942} &   \texttt{-202.3941} &   \texttt{-256.4609}&  \\
& (0.0060) &   (0.0070) &  (0.0110) &  (0.0100) &
\\
\hline
DMRG& -78.6857&  -154.8463& -202.5077 &-256.5535\\
\hline
\end{tabular}
\caption{VMC inference results for 2D TFIM Hamiltonian with various system sizes $(N_x, N_y)$  for $J=1$ and $B = 3.0$. These results were obtained from the inference process in which the trained model is used to generate new spin configurations whose mean energies are then calculated. For each ansatz, the mean $E$ is listed first, together with the corresponding standard error directly below (in brackets). The best results for each $(N_x, N_y)$ case are noted in typewriter font while the second best results are noted in bold. We emphasize that for this experiment, it is the second best results that we are interested in  (since the best result in each case is expected to be 2D Euclidean RNN)} \label{tab:2dtfim_res_int}
\end{table}
\FloatBarrier
From Table \ref{tab:2dtfim_res_int}, as expected, 2D Euclidean RNN is the best performing ansatz in all four cases, since the architecture of the RNN part of this NQS was specially created for the 2D setting of TFIM. Again, we stress that the goal of this experiment is not to determine the best, but rather, the second-best performing NQS. We are solely interested in determining the relative performance of 1D hyperbolic GRU with respect to 1D Euclidean GRU in the setting of 2D TFIM, with the inclusion of the 2D RNN NQS ansatz being made only to provide an additional benchmark. Regarding this aspect,  we observe that  hyperbolic GRU  outperformed Euclidean GRU 4 out of 4 times (see Fig.\ref{2d-tfim-comparison}). While for $(N_x, N_y) = (7,7), (8,8), (9,9)$, it is clear that hyperbolic GRU definitively outperformed Euclidean GRU, the situation is not clear-cut for $(N_x,N_y) = (5,5)$ when the performances are almost comparable, with hyperbolic GRU emerging as the slightly better ansatz.  
While more experiments involving larger $(N_x, N_y)$ sizes are needed, from the results of these experiments, it can be seen that 1D hyperbolic GRU would likely outperform 1D Euclidean GRU in 2D TFIM experiments.  
\\\\
This result is very interesting since it might be indicative of the role played by the hyperbolic geometry of the NQS ansatz in Hamiltonian systems when there exist some sort of hierarchical structure.
We recall that in the context of natural language processing (NLP) when the hyperbolic RNN/GRU was introduced in the work of \cite{ganea-1805}, the authors noted a definitive and clear outperformance of hyperbolic RNN/GRU over their Euclidean version when the underlying data exhibit a hierarchical, tree-like structure. In the present 2D TFIM setting, there exists a hierarchy in the interaction structure of 2D Hamiltonian when 1D NQS is used.
This is due to the artificial rearrangement (or folding) of the 1D spin chain samples generated by the 1D NQS to model the 2D data during the calculations of their local energies (see Fig.\ref{1d-to-2d} in Appendix \ref{sec-2d-rnn}). 
This process of rearrangement takes into account interactions between neighbors at site $i$ and site $i+N$, which were originally faraway in the 1D horizontal chain but are brought close together to form the vertical dimension in the 2D square lattice.
\\\\
Concretely, if we start with a two-dimensional $N\times N$ square lattice of spins whose sites are labeled as $(i',j')$ where $1\leq i'\leq N$ and $1\leq j'\leq N$, the 2D TFIM Hamiltonian of this system contains the nearest-neighbor interactions $\langle (i',j'), (i', j'+1)\rangle$ along the horizontal dimension as well as $\langle (i',j'), (i'+1, j')\rangle$ along the vertical dimension. When treated as a one-dimensional system, the 2D lattice can be thought of as a chain of length $N^2 = (N-1)N+N$  spins (see Subfigure (a) of Fig.\ref{1d-to-2d}) folded into a square lattice of size ($N,N$) (see Subfigure (b) of Fig.\ref{1d-to-2d}). The mapping of the site location in going from the 2D square spin lattice to the 1D spin chain system is:
\beq
\text{2D}: \,(i',j')\,\,\,\ra\,\,\text{1D}:\,\lf((i'-1)N + j'\rr)\,. \label{1d-2d}
\eeq
This leads to the original 2D nearest-neighbor interactions becoming the following 1D interactions:
\beq
\text{2D}:\,\lf\langle (i',j'), \frac{}{}(i', j'+1)\rr\rangle \,\,&\ra & \,\,\text{1D}: \,\lf\langle (i'-1)N + j', \frac{}{}(i'-1)N + j'+1 \rr\rangle\nonumber\\
\text{2D}:\,\lf\langle (i',j'),  \frac{}{}(i'+1, j')\rr\rangle \,\,&\ra & \,\,\text{1D}: \,\lf\langle (i'-1)N + j', \frac{}{} i'N + j' \rr\rangle
\eeq
Effectively, with a simple redefinition of the index $i = (i'-1)N+j'$, these 1D interactions are:
\beq
\lf\langle i, i+1\rr\rangle, \hspace{5mm} \langle i, i+N\rangle\,. \nonumber
\eeq
This means that in translating the 2D spin lattice to the 1D spin chain so that the 1D Euclidean and hyperbolic NQS ansatzes can be used (instead of the 2D Euclidean RNN ansatz), the original nearest-neighbor interactions along the horizontal and vertical dimensions of the 2D lattice become the nearest-neighbor $\langle i, i+1\rangle$ and the $N^{th}$-nearest neighbor $\langle i, i+N\rangle$ interactions in the 1D chain system.
A hierarchy of interactions is thus formed when both the the nearest and $N^{th}$-nearest neighbor interactions are present, and it is this hierarchy that might have played a part in the observation of 1D hyperbolic GRU outperforming 1D Euclidean GRU in this 2D TFIM setting.
\FloatBarrier
\subsection{1D Heisenberg $J_1J_2$ Model} \label{sec-res-j1j2}
The Hamiltonian of the one-dimensional $J_1J_2$ model is
\beq
H_{J_1-J_2} =  J_1 \sum_{\langle i, j\rangle} S_i\,.\,S_j + J_2 \sum_{\langle \langle i, j\rangle \rangle} S_i\,.\,S_j
\label{Hj1j2}
\eeq
The $S_i = S_x, S_y, S_z$ are the spin-half operators (Pauli matrices). The first sum with the coefficient $J_1$ in the Hamiltonian is over pairs of nearest neighbors while the second sum with the coefficient $J_2$ is over pairs of next nearest neighbors. When $J_2 = 0$, only nearest neighbor interactions are counted while next nearest neighbor interactions are switched off. It is known that from DMRG studies that this 1D model exhibits a phase transition at the critical $J_2$ value of $J_2^c = 0.2412 \pm 0.000 005$ from a critical Luttinger liquid phase ($J_2 \leq J_2^c)$ to a spontaneously gapped valence bond state phase ($J_2 \geq J_2^c)$ \cite{rnn_20}, \cite{j1j2-dmrg}, \cite{j1j2-pt1}, \cite{j1j2-pt2}. When $J_2 = 0.5$, the 1D $J_1J_2$ model reduce to the Majumdar-Ghosh model.
\subsubsection{Experiment details}
For the 1D $J_1J_2$ system with $J_1 = 1.0$ and $J_2 = 0.0, 0.2, 0.5, 0.8$, we carried out several VMC experiments using Euclidean and hyperbolic GRU NQS in the form of the complex NQS wavefunction Eq.(\ref{eq-wf-c}) described in \ref{sec-wf-c}. The architecture of these neural networks consist of a Euclidean/hyperbolic RNN unit and two Dense layers with 2 units each, one with the Softmax activation and one with the Softsign activation. 
\\\\
In all VMC experiments, the same Euclidean GRU variant, \texttt{eGRU-75-s50}, with the RNN hidden dimension $d_h = 75$ and 17854 trainable parameters\footnote{As a passing remark, we note that in the work \cite{rnn_20}, the authors used a much larger GRU-based model, comprising three layers of GRU, each with $d_h= 100$ to approximate the ground state energy of the 1D $J_1J_2$ model with $N=100$ spins. Their goal was to establish the performance of GRU-based neural network as an effective NQS. Here we are interested in establishing the performance of the hyperbolic version of the GRU-based NQS and not its Euclidean counterpart. This, and the fact that our 1D $J_1J_2$ system has a smaller number of spins $N$ ($N=50$ instead of 100), motivates our choice of using a smaller Euclidean GRU-based NQS.} is used to establish a benchmark performance against which we compare the performances of different variants of hyperbolic GRU. In particular, for each $J_2$ value, three variants of hyperbolic GRU with the RNN hidden dimension of $d_h = 60, 70, 75$ (\texttt{hGRU-60-s50}, \texttt{hGRU-70-s50}, \texttt{hGRU-75-s50}) were considered and used to run the VMC experiments, and the best result among these variants\footnote{Sometimes a variant with smaller $d_h$ hidden units outperforms one with larger $d_h$, as is the case for $J_2 = 0.0, 0.5$ when \texttt{hGRU-60-s50} and \texttt{hGRU-70-s50} outperform \texttt{hGRU-75-s50}, which ran into the problem of overfitting during the training process.} is selected to compare with the benchmark result provided by the Euclidean GRU. The list of all Euclidean/hyperbolic GRU variants is listed in Table \ref{tab:1dj1j2_setting}. 
\begin{table}[!ht]
\centering
\begin{tabular}{|l|c|c|c|c|}
\hline
Ansatz  & $J_2=0.0$ & $J_2=0.2$& $J_2=0.5$& $J_2=0.8$ \\
\hline
Euclidean GRU & \texttt{eGRU-75-s50} &\texttt{eGRU-75-s50}&\texttt{eGRU-75-s50} &\texttt{eGRU-75-s50} \\
Parameters & 17854 &  17854 &  17854 &  17854  \\
\hline
Hyperbolic GRU &\texttt{hGRU-60-s50} & \texttt{hGRU-75-s50} & \texttt{hGRU-70-s50}&\texttt{hGRU-75-s50} \\
 Parameters &  11584 & 17854 & 15614& 17854\\
\hline 
\end{tabular}
\caption{The  variants of Euclidean and hyperbolic GRU used as NQS ansatzes to run the VMC experiments involving the 1D $J_1J_2$ Hamiltonian Eq.(\ref{Hj1j2}) with $N=50$, $J_1 = 1$ and $J_2=0.0,0.2, 0.5, 0.8$. While three variants of hyperbolic GRU with $d_h=60,70,75$  were used in the experiments, we only list the best performing variant in this Table. The second and last rows titled `Parameters' list the numbers of parameters in the Euclidean and hyperbolic GRU networks.} \label{tab:1dj1j2_setting}
\end{table}
\FloatBarrier
Aside from the fact that the wavefunction in this case is complex instead of real, the other training details  remain exactly the same as described previously. 
The number of training epochs are 450 for all experiments.  Similar to previous cases, hyperbolic GRU takes 5-7 times longer to complete the training loop than Euclidean GRU due to their more involved mathematical construction\footnote{For $J_2 =0$, Euclidean GRU takes around 2.7 hours to complete the training, while for $J_2 \neq 0$, Euclidean GRU takes around 4 hours to complete the training on an average laptop with no GPU}.
\subsubsection{Results}\label{res-j1j2-2dg}
\begin{figure}[!h]
\centering
\includegraphics[width = .6\textwidth]{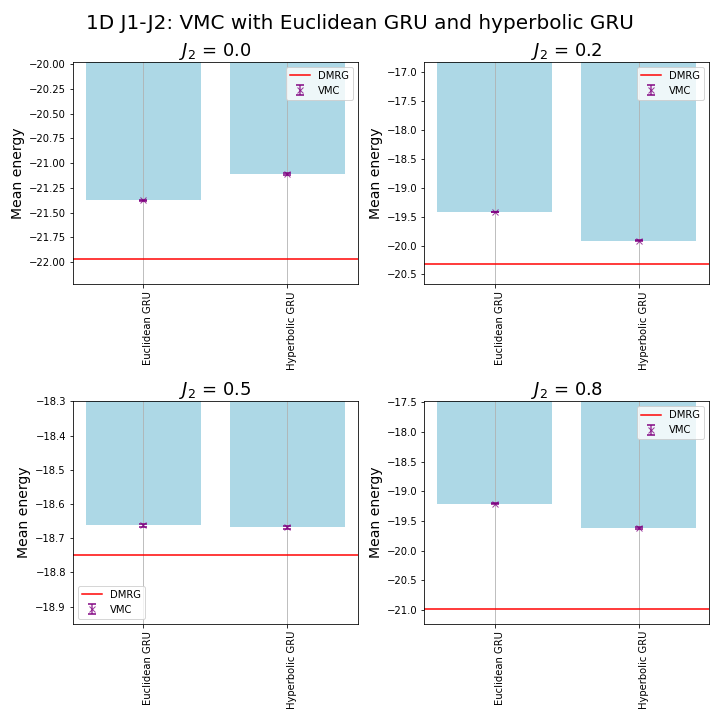}
\caption{Comparisons of the performances of Euclidean and hyperbolic GRU NQS ansatzes listed in Table \ref{tab:1dj1j2_setting} from VMC runs of 1D $J_1J_2$ model with different $J_2$ values (clockwise from top: $J_2$ =0.0, 0.2, 0.8, 0.5) for $N=50$ spins. In each subfigure, the mean energy for each NQS ansatz is shown as a dot with an error bar (representing the standard error). The specific variants of Euclidean/hyperbolic GRU used in each of the four cases are listed in Table \ref{tab:1dj1j2_setting}. In this plot, we do not specify the exact variants, but label them simply as Euclidean GRU or hyperbolic GRU to emphasize the geometry of the ansatzes.}\label{1d-j1j2-n50-comparison}
\end{figure}

In  Table \ref{tab:1dj1j2_res-gc} we list the results obtained by performing the inference process in which the trained models are used to generate $10^4$ new samples to calculate new NQS wavefunction. Fig.\ref{1d-j1j2-n50-comparison} graphically represents the data in Table \ref{tab:1dj1j2_res-gc} and shows the comparison of the performances between Euclidean and hyperbolic GRU NQS for all four values of $J_2$. In this case, unlike the TFIM cases, we note that the use of gradient clipping (both by value and by global norm\footnote{We performed the experiments using gradient clipping by both value and norm, and chose the best results among these.}) was necessary\footnote{The use of gradient clipping was necessary in almost all cases, with the exception of \texttt{hGRU-70-s50} in the case of $J_2 = 0.5$.  No gradient clipping was used during the training of hyperbolic GRU in that case (when $J_2 = 0.5$), as the energy curve for \texttt{hGRU-70-s50} showed no signs of any kinks or jumps.} to achieve stability (to avoid large jumps or kinks in the energy convergence curve) during the training process.  The convergence curve plots for the mean energy recorded during the training process with gradient clipping are shown in Fig.\ref{1d-j1j2-n50-gc}  in Appendix \ref{sec-app-j1j2}. 
\\\\
For completeness, we also included, in Appendix \ref{sec-app-j1j2}  the results obtained by training the NQS without using gradient clipping (see Table \ref{tab:1dj1j2_res-old} and Fig.\ref{1d-j1j2-n50-comparison-old} for the full results, see also Fig.\ref{1d-j1j2-n50} for the energy convergence graphs). Furthermore, Fig.\ref{1d-j1j2-n50-ernn-gc-ngc} and Fig.\ref{1d-j1j2-n50-hrnn-gc-ngc} show the energy convergence curves obtained during training with and without gradient clipping for Euclidean GRU and hyperbolic GRU, respectively.
\begin{table}[!h]
\centering
\begin{tabular}{lccccc} 
\hline\hline
Ansatz & $J_2=0.0$ & $J_2=0.2$& $J_2=0.5$& $J_2=0.8$  &\\
\hline\hline
Euclidean GRU
&  \textbf{-21.3745} & -19.4229  &  -18.6628 &   -19.2091 &
\\
&  (0.0060) &  (0.0039) &  (0.0053) &  (0.0132) &
\\
Hyperbolic GRU
&  -21.1108 & \textbf{-19.9133}   & \textbf{-18.6680}    &\textbf{-19.6191} &\\
&  (0.0110) &  (0.0065) &  (0.0047) &  (0.0144) &\\
\hline
Exact (DMRG)   & -21.9721& -20.3150 & -18.7500&  -20.9842 &\\
\hline
\end{tabular}
\caption{VMC inference results  (using 10000 samples) for the 1D $J_1J_2$ system with the Hamiltonian Eq.(\ref{Hj1j2}) with $N=50$, $J_1 = 1$ and $J_2=0.0,0.2, 0.5, 0.8$. For each ansatz, the mean $E$ is listed first, together with the corresponding standard error (in brackets). The best result at each $J_2$ value is noted in bold. } \label{tab:1dj1j2_res-gc}
\end{table}
\\
From  Table \ref{tab:1dj1j2_res-gc} and Fig.\ref{1d-j1j2-n50-comparison}, we note that hyperbolic GRU outperformed Euclidean GRU in three out of four cases. In particular, these three cases all correspond to nonzero values of $J_2$, i.e. when $J_2 =0.2, 0.5, 0.8$, in which the next nearest neighbor interactions $J_2$ are taken into account. When $J_2 = 0.2, 0.8$, hyperbolic GRU definitively outperformed Euclidean GRU, while for $J_2 = 0.5$, the two architectures showed almost comparable performances, with the hyperbolic GRU emerging as the marginally better performer. On the other hand, when $J_2 = 0.0$ (only nearest interactions $J_1$ - no next nearest neighbor interations $J_2$), Euclidean GRU outperformed hyperbolic GRU. It is noteworthy to point out that without gradient clipping, for $J_2\neq 0$, Euclidean GRU performed much worse than hyperbolic GRU (as is evident from Table \ref{tab:1dj1j2_res-old} and Fig.\ref{1d-j1j2-n50-comparison-old}). The use of gradient clipping eliminated the kinks/jumps in the energy convergence curves and improved the performances of the Euclidean GRU NQS significantly in the cases of $J_2 = 0.2$ and $J_2 = 0.5$ (see Fig.\ref{1d-j1j2-n50-ernn-gc-ngc} for a comparison of the energy convergence curves with and without gradient clipping). On the other hand, the use of gradient clipping only improved the performances of hyperbolic GRU NQS marginally, with the most notable improvements seen in the cases of $J_2 = 0.0$ and $J_2 = 0.8$ (see Fig.\ref{1d-j1j2-n50-hrnn-gc-ngc}). While gradient clipping improved the overall performances of both Euclidean and hyperbolic GRU (and much more so in the case of Euclidean GRU than hyperbolic GRU), the latter still ended up outperforming the former. 
\\\\
This result is again very interesting in the sense that it might be again indicative of the connection between the geometry of the NQS ansatz used and the structure of the Hamiltonian system under study, similar to the 2D TFIM setting above in which we observe a clear trend of outperformance by 1D hyperbolic GRU compared to 1D Euclidean GRU when there is a hierarchy in the Hamiltonian interaction structure, comprising the first and $N^{th}$ nearest neighbor interactions, introduced by the artificial rearrangement of the 1D spin chain to mimic the 2D spin lattice. 
 In this case, when second nearest neighbor interactions are switched on ($J_2 >0$), there is also a hierarchy going from the first nearest to the second nearest interactions, and it is our hypothesis that it is the decisive factor that leads to the outperformance of the hyperbolic GRU-based NQS over its Euclidean counterpart.
 \\\\
 It is unclear to us at the moment, although it is within our expectation, whether this would still be the case in two dimenions, i.e. whether hyperbolic GRU NQS carries on outperforming Euclidean GRU NQS in 2D $J_1J_2$ models with different 2D lattice geometries (such as square, triangle, kagome, etc.). In order to perform the VMC experiments for 2D $J_1J_2$ systems, one first needs to generalize the 1D hyperbolic GRU part of the NQS to two dimensions, similar to what was done in \cite{rnn_20} in generalizing the 1D Euclidean RNN cell to its 2D version in Eq.(\ref{eq-2d-rnn}). However, it is a highly nontrivial task to generalize to the 2D hyperbolic GRU, because even in Euclidean space, the mathematical construction of the 2D GRU is much more complex than that of the 1D GRU.  In one dimension, the hidden state $\vec h_i$ of the  1D GRU only depends on the previous state $\vec h_{i-1}$, resulting in the use of a single reset gate $\vec r$ and a single update gate $\vec z$ (see the GRU defining equations (\ref{gru-r})-(\ref{gru-z})), but in two dimensions, the hidden state $\vec h_{ij}$ of the 2D GRU receives updades from three different previous hidden states $\vec h_{i-1, j}$, $\vec h_{i,j-1}$ and $\vec h_{i-1, j-1}$, resulting in the use of three different reset gates $\vec r$'s and four different update gates $\vec z$'s (see Eq.\ref{eq:2d-gru-0} - Eq.\ref{eq:2d-gru-4}, which are the defining equations of the 2D Euclidean GRU as constructed in \cite{match-srnn}).\footnote{A concrete construction of the 2D GRU is given in the equations below from the work \cite{match-srnn}
\beq
\vec r_l &=& \s\lf(W_{r_l} \vec{q} + \vec b_{r_l}\rr), \hspace{5mm}
\vec r_t = \s\lf(W_{r_t} \vec{q} + \vec b_{r_t}\rr),\hspace{5mm}{}
\vec r_d = \s\lf(W_{r_d} \vec{q} + \vec b_{r_d}\rr), \label{eq:2d-gru-0}\\
\vec z_i &=& \s\lf(W_{z_i} \vec{q} + \vec b_{z_i}\rr), \hspace{5mm}
\vec z_l = \s\lf(W_{z_l} \vec{q} + \vec b_{z_l}\rr),\label{eq:2d-gru-1}\\
\vec z_t &=& \s\lf(W_{z_t} \vec{q} + \vec b_{z_t}\rr), \hspace{5mm}
\vec z_d =\s\lf(W_{z_d} \vec{q} + \vec b_{z_d}\rr),\label{eq:2d-gru-2}\\
\tilde{\vec{h}}_{ij} &=& f\lf(W\,\vec{x}_{ij} + U\lf[\vec{r} \oplus\lf\{\vec h_{i,j-1}, \vec h_{i-1,j}, \vec{h}_{i-1,j-1}\rr\}\rr] + \vec{b}\rr) \label{eq:2d-gru-3}\\
\vec{h}_{ij} &=& \vec{z}_l\oplus \vec{h}_{i,j-1} + \vec{z}_t \oplus \vec h_{i-1,j} \oplus \vec{z}_d\oplus \vec h_{i-1,j-1} + \vec z_i\oplus \tilde{\vec{h}}_{ij},
\label{eq:2d-gru-4}
\eeq
 where $\vec{q} = [\vec h_{i-1,j}, \vec h_{i,j-1}, \vec h_{i-1,j-1},\vec x_{ij}]^T$ is the concatenated vector containing the 3 previous hidden states $\vec h_{i-1,j}, \vec h_{i,j-1}, \vec h_{i-1,j-1}$ and the input vector $\vec x_{ij}$,  $\vec r_{l,t,d}$ are the three reset gates corresponding to $\vec h_{i, j-1}$, $\vec h_{i-1,j}$, $\vec h_{i-1,j-1}$, $\vec z_{l,t,d,i}$ are the four update gates. The $\vec W$'s, $\vec U$'s and $\vec b$'s are the weight matrices and bias vectors. $\s$ is the sigmoid function, $f$ is a nonlinear activation function, e.g. $\tanh$.}
 Even if we were to do away with the hidden state $\vec h_{i-1,j-1}$\footnote{The hidden state $\vec h_{i-1, j-1}$ might not be necessary in the context of 2D spin lattice with only vertical and horizontal nearest-neighbor interactions. In their construction of the 2D RNN, the authors of \cite{rnn_20} also did not make use of $\vec h_{i-1, j-1}$, see Eq.\ref{eq-2d-rnn}.}, we would still require two reset gates and three update gates for the 2D GRU. As such, with the many more parameters involved in the construction of the 2D Euclidean GRU and the already-complex definitions of various arithmetic operations in hyperbolic space, the hyperbolic 2D GRU (obtained from converting the 2D Euclidean GRU) will necessarily be much more mathematically complex than the 1D hyperbolic GRU. Such a network will be very computationally costly to train, both in terms of time and the hardware resources required.
\subsection{1D Heisenberg $J_1J_2J_3$ Model} \label{sec-res-j1j2j3}
The Hamiltonian of the one-dimensional Heisenberg $J_1J_2J_3$ model is 
\beq
H_{J_1J_2J_3} =  J_1 \sum_{\langle i, j\rangle} S_i\,.\,S_j + J_2 \sum_{\langle \langle i, j\rangle \rangle} S_i\,.\,S_j + J_3 \sum_{\langle \langle \langle i, j\rangle\rangle\rangle} S_i\,.\,S_j
\label{Hj1j2j3}
\eeq
where $S_i = S_x, S_y, S_z$ are the spin-half operators (Pauli matrices). The first sum with the coefficient $J_1$ in the Hamiltonian is over pairs of nearest neighbors, the second sum with the coefficient $J_2$ is over pairs of second nearest neighbors, and the third sum with the coefficient $J_3$ is over pairs of third nearest neighbors.
When both $J_2\neq 0$ and $J_3\neq 0$, this model has a complicated phase transition properties that depend almost entirely on the relative strength of the competing $J_2$ and $J_3$ interactions \cite{10-1d-j1j2j3}. Compared to the $J_1J_2$ model, the $J_1J_2J_3$ model is studied much less frequently in the literature, and even more so for the case of the one-dimensional $J_1J_2J_3$ model, but it is believed that the $J_3$ interactions in the 1D model between third nearest neighbors are relevant for describing the magnetic properties of quasi one-dimensional edge-shared cuprates such as LiVCuO$_4$ or LiCu$_2$O$_2$ \cite{10-1d-j1j2j3}. 

\subsubsection{Experiment details} 
For the 1D Heisenberg $J_1J_2J_3$ model with open boundary condition, we will carry out the VMC experiments with the number of spins $N=30$ using the following four settings (with $J_1=1.0$): 
\ben
\item $(J_2, J_3) = (0.0, 0.5)$: No second neighbor interactions, the $J_1J_2J_3$ model reduces to $J_1J_3$ model with only first and third nearest neighbor interactions.
 \item $(J_2, J_3) = (0.2, 0.2)$: Second and third neighbor interactions have the same strength. 
 \item $(J_2, J_3) = (0.2, 0.5)$: Second neighbor interactions are stronger than third neighbor interactions. 
 \item $(J_2, J_3) = (0.5, 0.2)$: Second neighbor interactions are weaker than third neighbor interactions. 
 \enn
The NQS used are the complex Euclidean GRU and hyperbolic GRU ansatzes as in the 1D $J_1J_2$ case. We will perform two different sets of VMC experiments. 
\bei
\item In the first set of experiments, all four ($J_2, J_3$) cases listed above are considered, with the Euclidean/hyperbolic GRU NQS ansatzes used having the same RNN hidden dimension of $d_h=50$ (see Table \ref{tab:1dj1j2j3_setting} for the variant names and their numbers of parameters).
\begin{table}[!ht]
\centering
\begin{tabular}{|l|c|c|}
\hline
($J_2, J_3)$ & Euclidean GRU  & Hyperbolic GRU \\
\hline
$\begin{matrix}
(0.0, 0.5)\\
(0.2, 0.2)\\
(0.2, 0.5)\\
(0.5, 0.2)\end{matrix}$ & $\begin{matrix}\texttt{eGRU-50-s50}\\(8154) \end{matrix}$ &$\begin{matrix}\texttt{hGRU-50-s50}\\(8154) \end{matrix}$ \\
\hline 
\end{tabular}
\caption{The  variants of Euclidean and hyperbolic GRU used as NQS ansatzes to run the first set of VMC experiments involving the 1D $J_1J_2J_3$ Hamiltonian Eq.(\ref{Hj1j2j3}) with $N=30$, $J_1 = 1$ and $(J_2, J_3) = (0.0, 0.5), (0.2, 0.2), (0.2, 0.5), (0.5, 0.2)$. The number of parameters of each NQS is listed in brackets under their names. Only one variant for each type of NQS ansatz is used in this set of VMC experiments. } \label{tab:1dj1j2j3_setting}
\end{table}
\item In the second set of experiments, we only considered $(J_2, J_3)$ = (0.2, 0.5), (0.5, 0.2), with the Euclidean GRU ansatz having the RNN hidden dimension $d_h = 60$. We experimented with various hyperbolic GRU ansatzes, each one having a different $d_h$ where $53 \leq d_h\leq 58$ in order to choose the best performing ansatz to compare the Euclidean NQS (see Table \ref{tab:1dj1j2j3_setting2}). 
\begin{table}[!ht]
\centering
\begin{tabular}{|l|c|c|}
\hline
($J_2, J_3)$ & Euclidean GRU  & Hyperbolic GRU \\
\hline
(0.2, 0.5) & $\begin{matrix}\texttt{eGRU-60-s50}\\(11584) \end{matrix}$ & $\begin{matrix}\texttt{hGRU-55-s50}\\(9794) \end{matrix}$\\
(0.5, 0.2)& $\begin{matrix}\texttt{eGRU-60-s50}\\(11584) \end{matrix}$ & $\begin{matrix}\texttt{hGRU-57-s50}\\(10492) \end{matrix}$ \\
\hline 
\end{tabular}
\caption{The  variants of Euclidean and hyperbolic GRU used as NQS ansatzes to run the second set of VMC experiments involving the 1D $J_1J_2J_3$ Hamiltonian Eq.(\ref{Hj1j2j3}) with $N=30$, $J_1 = 1$ and $(J_2, J_3) = (0.2, 0.5), (0.5, 0.2)$. Only one variant of Euclidean GRU, \texttt{eGRU-60-s50} with $d_h=60$ was used. Different variants of hyperbolic GRU with $ 53\leq d_h\leq 58$ were used in the experiment, but we only list the best performing variant in this Table. The number of parameters of each NQS is listed in brackets under their names. } \label{tab:1dj1j2j3_setting2}
\end{table}
\eni
 In the first set of VMC experiments, we only used one variant each for either Euclidean (\texttt{eGRU-50-s50} with $d_h=50$) or hyperbolic GRU (\texttt{hGRU-50-s50} with $d_h=50$) and fixed the number of training epochs to 280 in all cases. In the second set of VMC experiments, we explored the performances of hyperbolic GRU variants with $53\leq d_h \leq 58$ to choose the best one among them to compare with the Euclidean GRU benchmark, \texttt{eGRU-60-s50} with $d_h = 60$, while extending the number of training epochs to 450 for $(J_2, J_3) = (0.2, 0.5)$ and 500 for $(J_2, J_3) = (0.5, 0.2)$. 
 Similar to previous cases, in each experiment, hyperbolic GRU  can take up to around 5 times longer to complete the training (on the same hardware) than Euclidean GRU with the same $d_h$\footnote{while hyperbolic GRU with smaller $d_h$ also takes longer to train, up to around 2-3 times, than Euclidean GRU with larger $d_h$}. All other training details (optimizer settings, saving-of-best-model-only method) remain the same as in the 1D $J_1J_2$ case and will not be repeated here. 

\subsubsection{Results} 
\begin{figure}[!h]
\centering
\includegraphics[width = .6\textwidth]{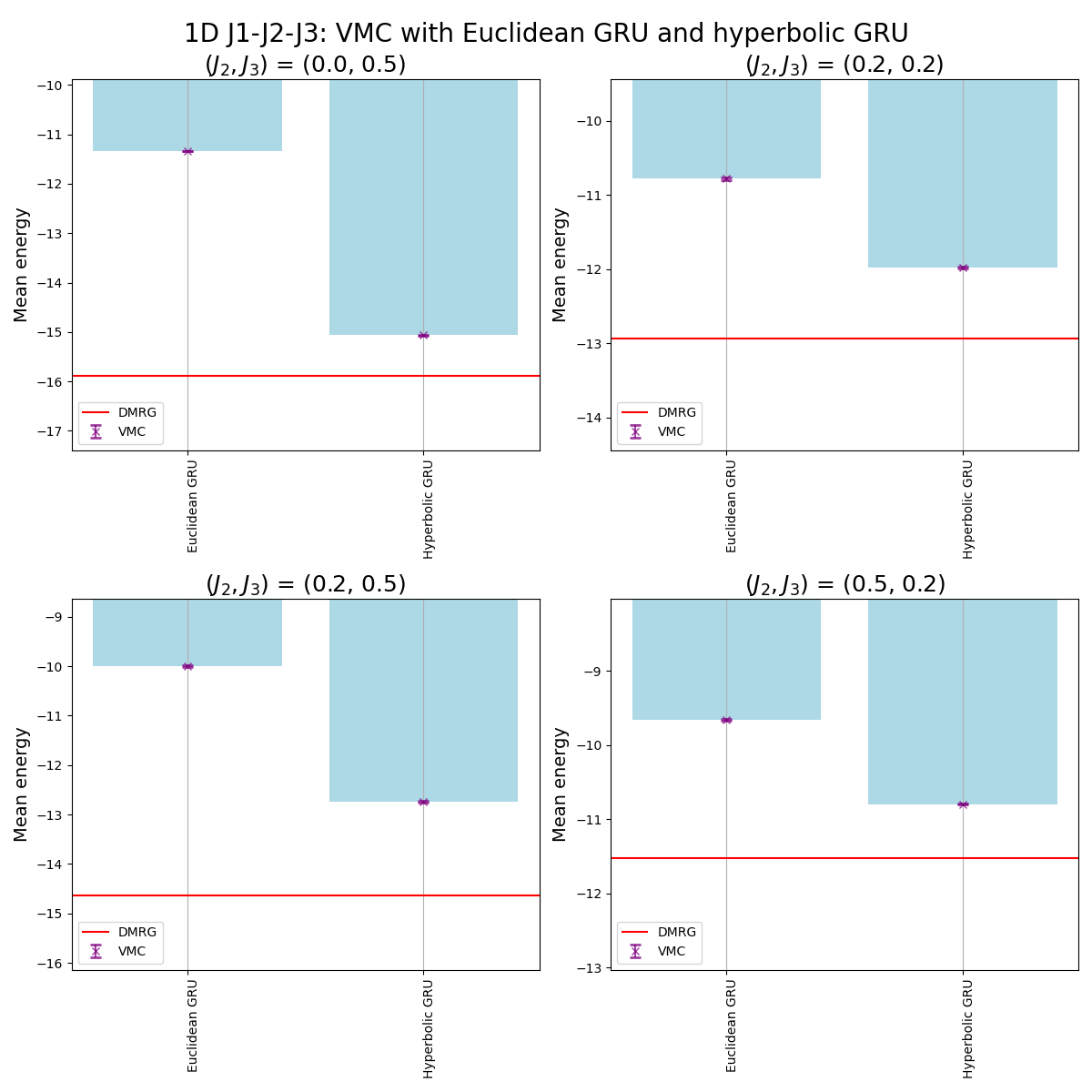}
\caption{Comparisons of the performances of Euclidean and hyperbolic GRU NQS ansatzes  listed in Table \ref{tab:1dj1j2j3_setting} from VMC runs of 1D $J_1J_2J_3$ model with different $(J_2,J_3)$ values - clockwise from top: (0.0, 0.5), (0.2, 0.2), (0.5, 0.2), (0.2, 0.5) - for $N=30$ spins. In each subfigure, the mean energy for each NQS ansatz is shown as a dot with an error bar (representing the standard error). As is the $J_1J_2$ case, we do not specify the exact variants (listed in Table \ref{tab:1dj1j2j3_setting}) in this plot, but choosing instead to label them as `Euclidean GRU' and `Hyperbolic GRU' to emphasize the geometry of the NQS ansatz.}\label{1d-j1j2j3-n30-comparison}
\end{figure}

\begin{figure}[!h]
\centering
\includegraphics[width = .6\textwidth]{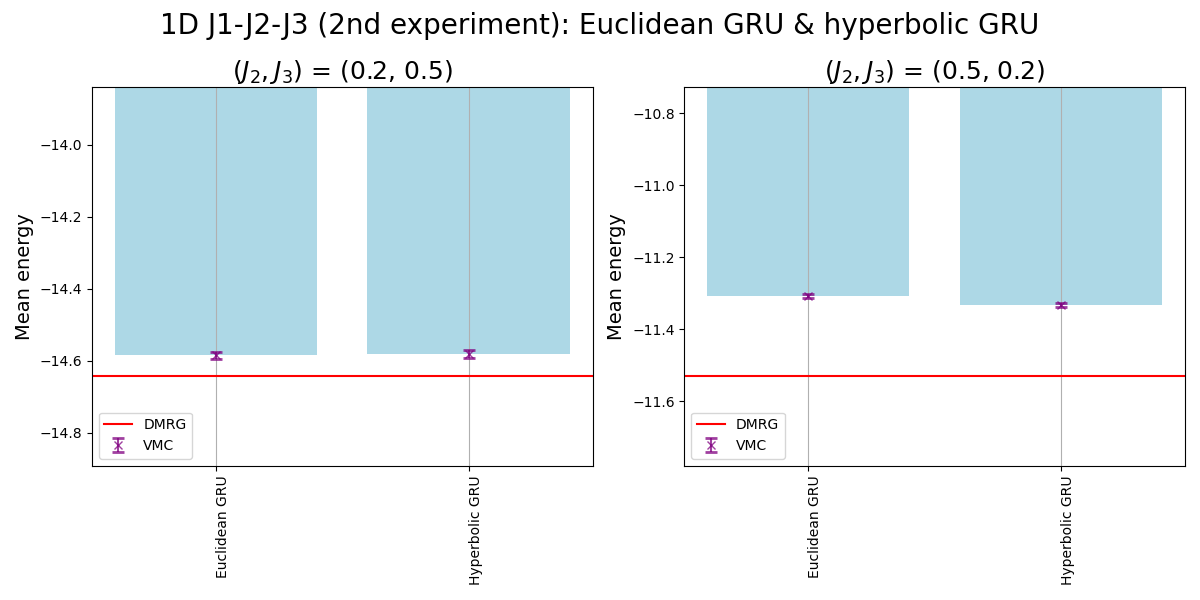}
\caption{Comparisons of the performances of Euclidean and hyperbolic GRU NQS ansatzes listed in Table \ref{tab:1dj1j2j3_setting2} from VMC runs of 1D $J_1J_2J_3$ model with different $(J_2,J_3)$ values - from left to right: (0.2, 0.5), (0.5, 0.2) - for $N=30$ spins. In each subfigure, the mean energy for each NQS ansatz is shown as a dot with an error bar (representing the standard error). As before, we do not specify the exact variants (listed in Table \ref{tab:1dj1j2j3_setting2}) in this plot, but choosing instead to label them as `Euclidean GRU' and `Hyperbolic GRU' to emphasize the geometry of the NQS ansatz.}\label{1d-j1j2j3-n30-comparison2}
\end{figure}

\bei
\item For the first set of VMC experiments involving $(J_2, J_3)$ = (0.0, 0.5), (0.2, 0.2), (0.2, 0.5), (0.5, 0.2) with the NQS ansatzes listed in Table \ref{tab:1dj1j2j3_setting}, the inference results using $10^4$ samples are listed in  Table \ref{tab:1dj1j2j3_int}, respectively. Fig.\ref{1d-j1j2j3-n30-comparison} graphically shows the data in both these Tables and offers a comparison between the performances of the 1D Euclidean and 1D hyperbolic GRU. 
The convergence curve plots for the mean energy recorded during the training process are shown in Fig.\ref{1d-j1j2j3-n30} in Appendix \ref{sec-app-j1j2j3}.
\begin{table}[!h]
\centering
\begin{tabular}{lccccc}
\hline\hline
Ansatz & $(J_2,J_3)=(0.0, 0.5)$ & $(J_2,J_3)=(0.2, 0.2)$& $(J_2,J_3)=(0.2, 0.5)$& $(J_2,J_3)=(0.5, 0.2)$ \\
\hline\hline
Euclidean GRU
&  -11.3393 & -10.7807   & -9.9980 & -9.6562& \\
&  (0.0070) &  (0.0150) &  (0.0240) &  (0.0140) &
\\
Hyperbolic GRU
&  \textbf{-15.0637} & \textbf{-11.9804} &    \textbf{-12.7448} & \textbf{-10.7973} &   \\
&   (0.0130) & (0.0150) &  (0.0180) &  (0.0060) &\\
\hline
Exact (DMRG) & -15.8903& -12.9430   & -14.6408   & -11.5287 &\\
\hline 
\end{tabular}
\caption{VMC inference results (from the first set of experiments with the NQS ansatzes listed in Table \ref{tab:1dj1j2j3_setting}) for the 1D $J_1J_2J_3$ system with the Hamiltonian Eq.(\ref{Hj1j2j3}) with $N=30$, $J_1 = 1$, and $(J_2,J_3) = (0.0, 0.2), (0.2, 0.2), (0.2, 0.5), (0.5, 0.2)$. For each ansatz, the mean $E$ is listed first, together with the standard error below (in brackets). The best result at each $(J_2, J_3)$ case is noted in bold. } \label{tab:1dj1j2j3_int}
\end{table}
\FloatBarrier
From Table \ref{tab:1dj1j2j3_int} and Fig.\ref{1d-j1j2j3-n30-comparison}, we note that hyperbolic GRU definitively outperformed Euclidean GRU (by a large margin) in all 4 out of 4 cases. The energy convergence curves (see Fig.\ref{1d-j1j2j3-n30}) corresponding to the hyperbolic GRU ansatz reach convergence at values well below that reached by the curves corresponding to the Euclidean GRU ansatz. 
\item
For the second set of VMC experiments involving only ($J_2, J_3) = (0.2, 0.5), (0.5, 0.2)$ with the NQS ansatzes listed in  Table \ref{tab:1dj1j2j3_setting2}, the inference results are included in Table \ref{tab:1dj1j2j3_int2}. The comparison of the performances of hyperbolic and Euclidean GRU ansatzes is shown in Fig.\ref{1d-j1j2j3-n30-comparison2}. The convergence curve plots of the second set of VMC experiments for the mean energy recorded during the training process are shown in Fig.\ref{1d-j1j2j3-n30_add} in Appendix \ref{sec-app-j1j2j3}.

\begin{table}[!h]
\centering
\begin{tabular}{lccccc}
\hline\hline
Ansatz & $(J_2,J_3)=(0.2, 0.5)$& $(J_2,J_3)=(0.5, 0.2)$ & \\
\hline\hline
Euclidean GRU
&       \textbf{-14.5846} &    -11.3073 & \\
&   (0.0100) & (0.0060) &
\\
Hyperbolic GRU
&    -14.5813   & \textbf{-11.3322} &\\
&  (0.0110) &  (0.0050) &\\
\hline
Exact (DMRG) &  -14.6408   & -11.5287 &\\
\hline 
\end{tabular}
\caption{VMC inference results (from the second set of experiments with the NQS ansatzes listed in Table \ref{tab:1dj1j2j3_setting2}) for the 1D $J_1J_2J_3$ system with the Hamiltonian Eq.(\ref{Hj1j2j3}) with $N=30$, $J_1 = 1$, and $(J_2,J_3) = (0.2, 0.5), (0.5, 0.2)$. For each ansatz, the mean $E$ is listed first, together with the corresponding standard error directly below (in brackets). The best result at each $(J_2, J_3)$ case is noted in bold. } \label{tab:1dj1j2j3_int2}
\end{table}
\ben
\item For $(J_2, J_3)$ = (0.2, 0.5), the hyperbolic GRU ansatz \texttt{hGRU-55-s50}, with 9794 parameters, shows an overall comparable performance to the Euclidean GRU ansatz \texttt{eGRU-60-s50} with 11584 parameters. Although the larger Euclidean GRU ansatz emerges as the better performer by a slight margin, it also has 20 percent more parameters than the hyperbolic GRU network. 
Furthermore, it is noteworthy to point out that in the training process, \texttt{hGRU-55-s50} performed better than \texttt{eGRU-60-s50}, as is evident from the energy convergence curve shown in the first subfigure of Fig.\ref{1d-j1j2j3-n30_add}, which shows the curve corresponding to \texttt{hGRU-55-s50} reaching convergence faster with a narrower fluctuation range than the curve corresponding to \texttt{eGRU-60-s50}. 
\item For $(J_2, J_3)$ = (0.5, 0.2), the hyperbolic GRU ansatz \texttt{hGRU-57-s50} outperforms the Euclidean GRU ansatz \texttt{eGRU-60-s50}. In this case, hyperbolic GRU has 10 percent fewer parameters than Euclidean GRU (10492 versus 11584). 
\enn
\eni
From the two sets of VMC experiments for the 1D $J_1J_2J_3$ model, we observe that the obtained results showing that hyperbolic GRU almost always outperformed Euclidean GRU are in agreement with the ones in the 1D $J_1J_2$ case in which hyperbolic GRU outperfomed its Euclidean version in all cases when $J_2\neq 0$, i.e. when the second neighbor interaction is nonvanishing and the Hamiltonian exhibits a hierarchy in the interaction structure comprised of the first and second neighbor interactions. 
In all four $J_1J_2J_3$ settings where there exist an even more pronounced hierarchy in the interaction structure of the Hamiltonian that comprises the first, second and third nearest neighbor interactions, the clear outperformance of the hyperbolic GRU ansatz reinforces the hypothesis that the hyperbolic geometry of this neural network plays a determining role in handling the type of Hamiltonians with hierarchical structures such as the 1D $J_1J_2$, $J_1J_2J_3$ systems. As already noted in the 1D $J_1J_2$ case (see Section \ref{res-j1j2-2dg}), an interesting task would be to verify whether hyperbolic GRU carries on outperforming Euclidean GRU for the 2D $J_1J_2J_3$ models on different lattice geometries, but this will require the generalization of the 1D hyperbolic GRU network to a 2D version, similar to the generalization of the 1D Euclidean RNN cell to the 2D version in Eq.\ref{eq-2d-rnn} as done in \cite{rnn_20}. As also noted in Section \ref{res-j1j2-2dg}, this generalization is a challenging task, given that the complexity of the resulting 2D hyperbolic network will result in training processes that are highly computationally demanding.
\section{Concluding remarks} \label{sec-concl}
In this work, we introduce the hyperbolic GRU as the first type of non-Euclidean neural quantum state (NQS) to be used in the variational Monte Carlo method (VMC) to approximate the ground state energy in quantum many-body physics, which are represented by the prototypical settings of the one-dimensional \& two-dimensional transverse field Ising model (1D/2D TFIM) and the one-dimensional Heisenberg $J_1J_2$ \& $J_1J_2J_3$ models. Our results show an interesting trend regarding the performance of the hyperbolic GRU as benchmarked against the performances of its Euclidean counterpart. In particular, from the obtained results of the experiments performed in this work, we note the following.
\bei
\item In the 1D TFIM setting, with the number of spins $N$ being 20, 40, 80 and 100, where the Hamiltonian comprises only the nearest neighbor interaction and a transverse magnetic field, with no hierarchical interaction structure, hyperbolic GRU showed a comparable performance to Euclidean GRU, with both GRU variants outperforming the Euclidean RNN variant. 
\item In the 2D TFIM setting, with the number of size of the 2D square lattice  $(N_x, N_y)$ being $(5,5), (7,7), (8,8)$, $(9,9)$ corresponding to $N=25,49, 64, 81$ spins, where the Hamitonian comprises the horizontal and vertical nearest neighbor interaction in two dimensions and a transverse magnetic field, 1D hyperbolic GRU always outperformed 1D Euclidean GRU (while the best performing ansatz, the 2D Euclidean RNN NQS, was included purely to provide a benchmark). In this case, due to the artificial rearrangement or reshaping/folding of the 1D spin chain to mimic the 2D lattice, a hierarchy in the interaction structure, comprising the first and the $N^{th}$ nearest neighbor interactions, is introduced when originally faraway spins (at site $i$ and site $i+N$) in the 1D chains become immediate vertical neighbors. The fact that the hyperbolic GRU NQS ansatz outperformed the Euclidean GRU NQS ansatz for this setting is our first clue of the role played by the hyperbolic geometry of the neural network in determining its performance when there is a hierarchical structure in the Hamiltonian. 
\item In the 1D  Heisenberg $J_1J_2$ setting with $N=50$ spins, $J_1 =1.0$ (fixed) and $J_2 = 0.0, 0.2, 0.5, 0.8$, hyperbolic GRU definitively outperformed Euclidean GRU when the $J_1J_2$ Hamiltonian exhibits a hierarchical interaction structure - when $J_2 \neq 0$, which takes into account the second nearest neighbor interactions in addition to the first nearest neighbor interactions. This is our second clue of the role played by the hyperbolic geometry of the neural network.
\item In the 1D Heisenberg $J_1J_2J_3$ setting with $N=30$ spins, $J_1 = 1.0$ (fixed) and $(J_2, J_3)=(0.0, 0.5), (0.2,0.2)$, $(0.2, 0.5)$, $(0.5, 0.2)$, hyperbolic GRU again definitively outperformed Euclidean GRU in four out of four cases in the first set of experiments (where the networks have the same number of parameters). In the second set of experiments (involving only the two cases $J_2 = (0.2, 0.5)$ and $J_2 = (0.5, 0.2)$), where the hyperbolic GRU network has around 10-20 percent fewer parameters than the Euclidean GRU network, hyperbolic GRU attained a comparable performance with Euclidean GRU in one case and outperformed Euclidean GRU in another case. In this setting, the $J_1J_2J_3$ Hamiltonian  exhibits even more pronounced hierarchical interaction structures comprising the first, second and third nearest neighbors. This is our third clue. 
\eni
At this point, it is useful to recall the established results in natural language processing (NLP) and graph embedding that hyperbolic RNNs almost always outperform their Euclidean versions when the training data exhibit hierarchical structures \cite{ganea-1805}, \cite{2010-hyprnn}, \cite{hyprnn-review-2101}. In these NLP/graph embedding contexts, the fact that hierarchical structures are often tree-like in nature largely explains the superior performances of hyperbolic neural networks, since a key property of hyperbolic space - the exponential expansion of space at large distances - is responsible for its ability to accomodate or embed tree structures with very low distortions. In the context of physical Hamiltonian spin systems, the hierarchical structures are not really tree-like in nature,  but nonetheless a clear hierarchy in the interaction part exists in all the systems that hyperbolic NQS was found to have outperformed its Euclidean counterpart. Thus, it is  our hypothesis that hyperbolic RNN-based NQS will \textit{likely} outperform Euclidean RNN-based NQS in all quantum spin systems in which there is a hierarchical structure in the interaction part of the Hamiltonian. Given the fact that the experiments done in this work have only involved small-size systems due to our limited computing resources, it is obvious that more works are needed to either confirm or refute this hypothesis. If true, this would be an exciting new direction opened up by the possibility of better performances delivered hyperbolic and potentially other non-Euclidean NQS in the problem of approximating the ground state energy of quantum many-body Hamiltonians using VMC - a problem that has been studied using exclusively conventional or Euclidean neural networks.
\\\\
\noindent Several interesting generalizations exist given the results of this work.
\bei
\item One immediate and specific direction involves the generalization of the one-dimensional hyperbolic GRU NQS in this work to two dimensions, and compare the performance of this 2D hyperbolic GRU against the 2D Euclidean GRU in the settings of 2D TFIM or 2D Heisenberg $J_1J_2$ and $J_1J_2J_3$ models. The main challenge for this task lies in the fact that the computational cost required to train the 2D Euclidean and hyperbolic GRUs will be much more intensive than the already-intensive computational cost required to train the 1D hyperbolic GRU. This is because given the fact that the 2D Euclidean GRU is more mathematically complex than the 1D GRU, its hyperbolic conversion will necessarily result in an even more complex{} construction, as previously noted in detail in Section \ref{res-j1j2-2dg}.

\item Other broader directions involve the exploration of a different type of hyperbolic RNNs, for example one that is based on the Lorentz model of hyperbolic space  \cite{hyprnn-lorentz-1806}, \cite{hyprnn-lorentz-2105} instead of the Poincar\'e model, in various quantum many-body settings. It is known that hyperbolic networks based on the Lorentz model exhibit a more stable behavior compared to  those based on the Poincar\'e model during the training process\footnote{In the Poincar\'e model, the metric includes a term $2/(1-||x||^2)$ (see Eq.\ref{eq:poincare-metric}), which causes a `singularity' at the boundary, or in other words, the distance grows to infinity at the edge of the Poincar\'e ball. During the training process, a small update in the vicinity of the boundary can lead to a very large change in the hyperbolic distance and subsequently a gradient explosion problem. On the other hand, the Lorentz model of $D$-dimensional hyperbolic space does not have a singularity in its metric, since it is defined as a $D$-dimensional hypersurface $\mathcal H^D$ in $(D+1)$ dimensional Euclidean space $\mathbb R^{D+1}$:
\beq
\mathcal H^D = \lf\{\vec x \in \mathbb R^{D+1}: \langle \vec x, \vec x\rangle =-1, x_0>0\rr\}, \hspace{4mm} \text{where}\hspace{3mm}\langle \vec x, \vec y \rangle = -x_0y_0 +\sum_{i=1}^D x_iy_i\,.
\eeq
The distance function in $\mathcal H^D$ is $d(\vec x, \vec y) = \text{arccosh}\lf(-\langle \vec x, \vec y\rangle\rr)$, which does not have a singularity.
 },
 and this is especially the case with larger or deeper hyperbolic networks \cite{hyprnn-lorentz-1806}. With the increasing sizes of the quantum systems of interest, it might be more suitable to switch from Poincar\'e-based hyperbolic networks to the Lorentz-based hyperbolic ones for precisely this reason.
\item Yet even more general directions involve the explorations of different types of non-Euclidean NQS beyond hyperbolic RNNs. These new types of non-Euclidean NQS could include, for example, the hyperbolic CNN networks \cite{2006-hypcnn} or the hyperbolic transformer neural network \cite{hyprnn-lorentz-2105}.
\eni
 We leave these issues to future works.
\appendix
\section{Two-dimensional RNN model} \label{sec-2d-rnn}

The 2D Euclidean RNN is a new custom construction from \cite{rnn_20} with the following defining equation\footnote{A more general discussion of the multi-dimensional RNN can be found in \cite{md-rnn-07}.}
 \beq
 \vec h_{t+1} = f\lf(U_h \vec x^0_t + W_h \vec h^0_t +  U_v \vec x^1_t + W_v \vec h_t^0 + \vec b\rr)\,, \label{eq-2d-rnn}
 \eeq
 where $f$ is a nonlinear activation function, the input $\vec x_t=(\vec x^0_t, \vec x^1_t)$ and hidden state $\vec h_t =(\vec h^0_t, \vec h^1_t)$ at timestep $t$ are now a 2D vectors comprising the horizontal and vertical components. Correspondingly, there is a doubling of the ususal weight matrices $U$ and $W$ to $U_h, U_v$, $W_h, W_v$. For a 2D spin configuration $\mathbf{\s}$ of dimension $(N_x, N_y$), an entry $\s_{i,j}$ is autoregressively generated based on the two spins and two hidden states from its previously generated horizontal and vertical neighbors. 
For the 1D Euclidean/hyperbolic GRU, during the sampling process, the site $\s_{i,j}$ is autoregressively generated based on only the previously generated site in a raster manner shown in Fig.\ref{2dtfim_sampling}. At each time step $t$ during the training phase, the 2D Euclidean RNN NQS ansatz generates 50 two-dimensional samples, each of which is a (2D) spin configuration (samples) of size $(N_x, N_y$), while the 1D Euclidean/hyperbolic GRU ansatz generates 50 one-dimensional samples, each of which is a (1D) spin configuration of length $N_x\times N_y$. The computation of the energy for the 1D ansatz requires a reshaping of the 1D samples of length $N_x\times N_y$ into 2D samples of size $(N_x, N_y)$. 
\begin{figure}[H]
\centering
\begin{subfigure}[b]{0.45\textwidth}
\centering
\begin{tikzpicture}[node distance = 1.3cm, thick]%
        \node[circle, draw] (1) {$\,1\,$};
        \node[circle, draw] (2) [right of =1]{$\,2\,$};
        \node[circle, draw] (3) [right of =2]{$\,3\,$};
        \node[circle, draw] (4) [right of =3]{$\,4\,$};
        \node[circle, draw] (5) [below of =1]{$\,5\,$};
        \node[circle, draw] (6) [right of =5]{$\,6\,$};
        \node[circle, draw] (7) [right of =6]{$\,7\,$};
        \node[circle, draw] (8) [right of =7]{$\,8\,$};
        \node[circle, draw] (9) [below of =5]{$\,9\,$};
        \node[circle, draw] (10) [right of =9]{$10$};
        \node[circle, draw] (11) [right of =10]{$11$};
        \node[circle, draw] (12) [right of =11]{$12$};
        \node[circle, draw] (13) [below of =9]{$13$};
        \node[circle, draw] (14) [right of =13]{$14$};
        \node[circle, draw] (15) [right of =14]{$15$};
        \node[circle, draw] (16) [right of =15]{$16$};
        \draw[->] (1) -- node [right]{} (2);
        \draw[->] (2) -- node [right]{} (3);
        \draw[->] (3) -- node [right]{} (4);
        \draw[->] (4) -- node [right]{} (5);
        \draw[->] (5) -- node [right]{} (6);
        \draw[->] (6) -- node [right]{} (7);
        \draw[->] (7) -- node [right]{} (8);
        \draw[->] (8) -- node [right]{} (9);
        \draw[->] (9) -- node [right]{} (10);
        \draw[->] (10) -- node [right]{} (11);
        \draw[->] (11) -- node [right]{} (12);
        \draw[->] (12) -- node [right]{} (13);
        \draw[->] (13) -- node [right]{} (14);
        \draw[->] (14) -- node [right]{} (15);
        \draw[->] (15) -- node [right]{} (16);
    \end{tikzpicture}
    \caption{}
\end{subfigure}
\begin{subfigure}[b]{0.45\textwidth}
\centering
\begin{tikzpicture}[node distance = 1.3cm, thick]%
        \node[circle, draw] (1) {$\,1\,$};
        \node[circle, draw] (2) [right of =1]{$\,2\,$};
        \node[circle, draw] (3) [right of =2]{$\,3\,$};
        \node[circle, draw] (4) [right of =3]{$\,4\,$};
        \node[circle, draw] (8) [below of =1]{$\,8\,$};
        \node[circle, draw] (7) [right of =8]{$\,7\,$};
        \node[circle, draw] (6) [right of =7]{$\,6\,$};
        \node[circle, draw] (5) [right of =6]{$\,5\,$};
        \node[circle, draw] (9) [below of =8]{$\,9\,$};
        \node[circle, draw] (10) [right of =9]{$10$};
        \node[circle, draw] (11) [right of =10]{$11$};
        \node[circle, draw] (12) [right of =11]{$12$};
        \node[circle, draw] (13) [below of =12]{$13$};
        \node[circle, draw] (14) [left of =13]{$14$};
        \node[circle, draw] (15) [left of =14]{$15$};
        \node[circle, draw] (16) [left of =15]{$16$};
        \draw[->] (1) -- node [right]{} (2);
        \draw (1) -- node []{} (8);
        \draw[->] (2) -- node [right]{} (3);
        \draw (2) -- node []{} (7);
        \draw[->] (3) -- node [right]{} (4);
        \draw (3) -- node []{} (6);
        \draw[->] (4) -- node [right]{} (5);
        \draw[->] (5) -- node [right]{} (6);
        \draw (5) -- node []{} (12);
        \draw[->] (6) -- node [right]{} (7);
        \draw (6) -- node []{} (11);
        \draw[->] (7) -- node [right]{} (8);
        \draw (7) -- node []{} (10);
        \draw[->] (8) -- node [right]{} (9);
        \draw[->] (9) -- node [right]{} (10);
        \draw (9) -- node []{} (16);
        \draw[->] (10) -- node [right]{} (11);
        \draw (10) -- node []{} (15);
        \draw[->] (11) -- node [right]{} (12);
        \draw (11) -- node []{} (14);
        \draw[->] (12) -- node [right]{} (13);
        \draw[->] (13) -- node [right]{} (14);
        \draw[->] (14) -- node [right]{} (15);
        \draw[->] (15) -- node [right]{} (16);
    \end{tikzpicture}
    \caption{}
    \end{subfigure}
    \caption{Schematic of the sampling process in the 2D TFIM model with a square lattice of size $(N_x, N_y) = (4,4)$ using (a) 1D Euclidean/hyperbolic GRU (b) 2D Euclidean RNN. The arrows in both (a) and (b) denote the autoregressive sampling path. For 2D RNN, each site in the square lattice receives 2 RNN hidden states from its nearest horizontal and vertical neighbors, while for 1D GRU, each site only receives the hidden state from its original nearest neighbor in the 1D chain. Figure adapted from \cite{rnn_20}.} \label{2dtfim_sampling}
    \end{figure}

\FloatBarrier

\begin{figure}[H]
\centering
\begin{subfigure}[b]{0.9\textwidth}
\centering
\begin{tikzpicture}[node distance = 1.3cm, thick]%
        \node[circle, draw] (1) {$\,1\,$};
        \node[circle, draw] (2) [right of =1,xshift = 0.5cm]{$\,2\,$};   
        \node[rectangle, draw] (3) [right of =2,xshift = 0.5cm]{$\,\ldots\,$};
        \node[rectangle, draw] (4) [right of =3,xshift = 0.5cm]{$\,\,N\,\,$};
        \node[rectangle, draw] (5) [right of =4, xshift = 0.5cm]{$\,N+1\,$};
        \node[rectangle, draw] (6) [right of =5, xshift = 0.5cm]{$\,\ldots\,$};
        \node[rectangle, draw] (7) [right of =6, xshift = 0.5cm]{$\,N+N\,$};
        \node[rectangle, draw] (8) [right of =7, xshift = 0.5cm]{$\,\ldots\,$};
        \node[rectangle, draw] (9) [right of =8, xshift = 1.0cm]{$(N-1)N+N$};
        \draw[-] (1) -- node [right]{} (2);
        \draw[-] (2) -- node [right]{} (3);
        \draw[-] (3) -- node [right]{} (4);
        \draw[-] (4) -- node [right]{} (5);
        \draw[-] (5) -- node [right]{} (6);
        \draw[-] (6) -- node [right]{} (7);
        \draw[-] (7) -- node [right]{} (8);
        \draw[-] (8) -- node [right]{} (9);       
    \end{tikzpicture}
    \caption{}
\end{subfigure}
\begin{subfigure}[b]{0.7\textwidth}
\centering
\begin{tikzpicture}[node distance = 1.3cm, thick]%
        \node[circle, draw] (1) {$\,1\,$};
        \node[circle, draw] (2) [right of =1, xshift = 1.5cm]{$\,2\,$};
        \node[circle, draw] (3) [right of =2,xshift = 1.5cm]{$\,3\,$};
        \node[rectangle, draw] (4) [right of =3,xshift = 1.5cm]{$\,\ldots\,$};
        \node[circle, draw] (4b) [right of =4,xshift = 1.5cm]{$\,N\,$};

        \node[rectangle, draw] (5) [below of =1]{$\,N+1\,$};
        \node[rectangle, draw] (6) [below of =2]{$\,N+2\,$};
        \node[rectangle, draw] (7) [below of =3]{$\,N+3\,$};
        \node[rectangle, draw] (8) [below of =4]{$\,\ldots\,$};
        \node[rectangle, draw] (8b) [below of =4b]{$\,N+N\,$};

        \node[rectangle, draw] (9) [below of =5]{$\,2N+1\,$};
        \node[rectangle, draw] (10) [below of =6]{$2N+2$};
        \node[rectangle, draw] (11) [below of =7]{$2N+3$};
        \node[rectangle, draw] (12) [below of =8]{$\,\ldots\,$};
        \node[rectangle, draw] (12b) [below of =8b]{$2N+N$};

        \node[rectangle, draw] (13) [below of =9]{$\,\ldots\,$};
        \node[rectangle, draw] (14) [below of =10]{$\,\ldots\,$};
        \node[rectangle, draw] (15) [below of =11]{$\,\ldots\,$};
        \node[rectangle, draw] (16) [below of =12]{$\,\ldots\,$};
        \node[rectangle, draw] (16b) [below of =12b]{$\,\ldots\,$};

        \node[rectangle, draw] (17) [below of =13]{$(N-1)N+1$};
        \node[rectangle, draw] (18) [below of =14]{$(N-1)N+2$};
        \node[rectangle, draw] (19) [below of =15]{$(N-1)N+3$};
        \node[rectangle, draw] (20) [below of =16]{$\,\ldots\,$};
        \node[rectangle, draw] (20b) [below of =16b]{$(N-1)N+N$};
        \draw[-] (1) -- node [right]{} (2);
        \draw[-] (2) -- node [right]{} (3);
        \draw[-] (3) -- node [right]{} (4);
        \draw[-] (4) -- node [right]{} (4b);
        \draw[-] (5) -- node [right]{} (6);
        \draw[-] (6) -- node [right]{} (7);
        \draw[-] (7) -- node [right]{} (8);
        \draw[-] (8) -- node [right]{} (8b);
        \draw[-] (9) -- node [right]{} (10);
        \draw[-] (10) -- node [right]{} (11);
        \draw[-] (11) -- node [right]{} (12);
        \draw[-] (12) -- node [right]{} (12b);
        \draw[-] (13) -- node [right]{} (14);
        \draw[-] (14) -- node [right]{} (15);
        \draw[-] (15) -- node [right]{} (16);
        \draw[-] (16) -- node [right]{} (16b);
        \draw[-] (17) -- node [right]{} (18);
        \draw[-] (18) -- node [right]{} (19);
        \draw[-] (19) -- node [right]{} (20);
        \draw[-] (20) -- node [right]{} (20b);
        \draw[-] (5) -- node [right]{} (1);
        \draw[-] (6) -- node [right]{} (2);
        \draw[-] (7) -- node [right]{} (3);
        \draw[-] (8) -- node [right]{} (4);
        \draw[-] (8b) -- node [right]{} (4b);
        \draw[-] (9) -- node [right]{} (5);
        \draw[-] (10) -- node [right]{} (6);
        \draw[-] (11) -- node [right]{} (7);
        \draw[-] (12) -- node [right]{} (8);
        \draw[-] (12b) -- node [right]{} (8b);
        \draw[-] (13) -- node [right]{} (9);
        \draw[-] (14) -- node [right]{} (10);
        \draw[-] (15) -- node [right]{} (11);
        \draw[-] (16) -- node [right]{} (12);
        \draw[-] (16b) -- node [right]{} (12b);
        \draw[-] (17) -- node [right]{} (13);
        \draw[-] (18) -- node [right]{} (14);
        \draw[-] (19) -- node [right]{} (15);
        \draw[-] (20) -- node [right]{} (16);
        \draw[-] (20b) -- node [right]{} (16b);
    \end{tikzpicture}
    \caption{}
\end{subfigure}
\caption{The process of rearranging or reshaping a 1D chain of length $N^2 = (N-1)N+N$ into a 2D lattice $N\times N$ introduces additional interactions between site $i$ and site $(i+N)$ when the vertical dimension is `formed'. In the original 1D lattice in the setting of 1D TFIM, there exist only the nearest-neighbor interactions $\langle i, i+1\rangle$. When the 1D spin chain (with site locations labeled bya single index $i$)is reshaped into a 2D square lattice (with site locations labeled by $(i',j')$) in the setting of 2D TFIM, besides the 1D original $\langle i,i+1\rangle$ interaction which corresponds to the 2D nearest neighbor interaction $\langle (i',j'), (i',j'+1)\rangle$ along the horizontal axis, an additional type of interactions $\langle i, i+N\rangle$, which corresponds to the 2D vertical nearest neighbor interaction $\langle (i',j'), (i'+1, j')\rangle$, is introduced (see Eq.\ref{1d-2d} for the 1D-2D mapping and the discussion in Section \ref{sec-res-2dtfim-res} for more details). (a) The original 1D chain with length $N^2 = (N-1)N+N$, with each site in the chain labeled by the 1D index $1\leq i \leq (N-1)N+N$. (b) The same 1D chain of length $N^2 = (N-1)N+N$ rearranged in a 2D square lattice of size $N\times N$, with each site in the square lattice labeled by their original 1D position. } \label{1d-to-2d}
\end{figure}

\FloatBarrier
\newpage
\section{Convergence curves}
\subsection{1D TFIM}\label{sec-app-1dtfim}
\begin{figure*}[ht!]
    \centering
    \includegraphics[width = .65\textwidth]{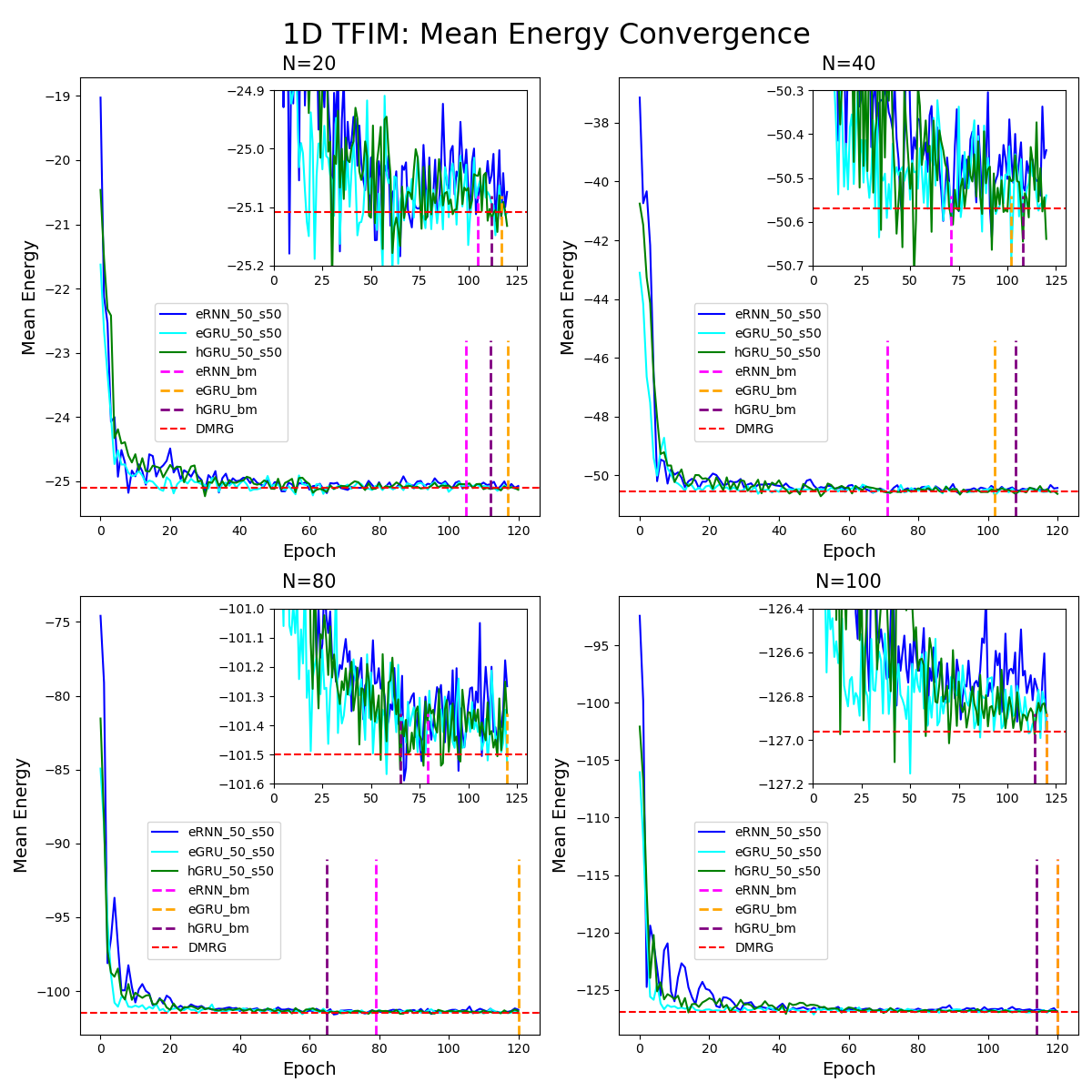}
\caption{Convergence curves of the mean energy from the VMC experiments for 1D TFIM with various $N$ (clockwise from top: $N =20,40, 100, 80$ spins) with the three types of NQS ansatzes being \texttt{eRNN-50-s50}, \texttt{eGRU-50-s50}, \texttt{hypGRU-50-s50}. Each subfigure contains a zoomed-in plot showing a narrower range of energy values close to the exact DMRG value. The vertical dashed lines (whose legend labels contain the letters `\texttt{bm}') indicate the epochs at which the best models were saved. The epochs at which the best models were saved were automatically determined by the training loop such that after convergence is achieved, the best models not only have the lowest mean energy but also with the variance below the tolerance threshold (see the discussion in Section \ref{sec-res} just before Section 5.1). When $N=100$, the best models for \texttt{eGRU} and \texttt{eRNN} were both saved at epoch 120, thus only one vertical dashed line is visible, since these 2 dashed lines corresponding to  \texttt{eGRU} and \texttt{eRNN} coincide with each other. }
\label{fig_1dtfim_n20_40_80}
\end{figure*}

\FloatBarrier
\newpage
\subsection{2D TFIM}\label{sec-app-2dtfim}
\begin{figure}[!h]
\centering
\includegraphics[width = .67\textwidth]{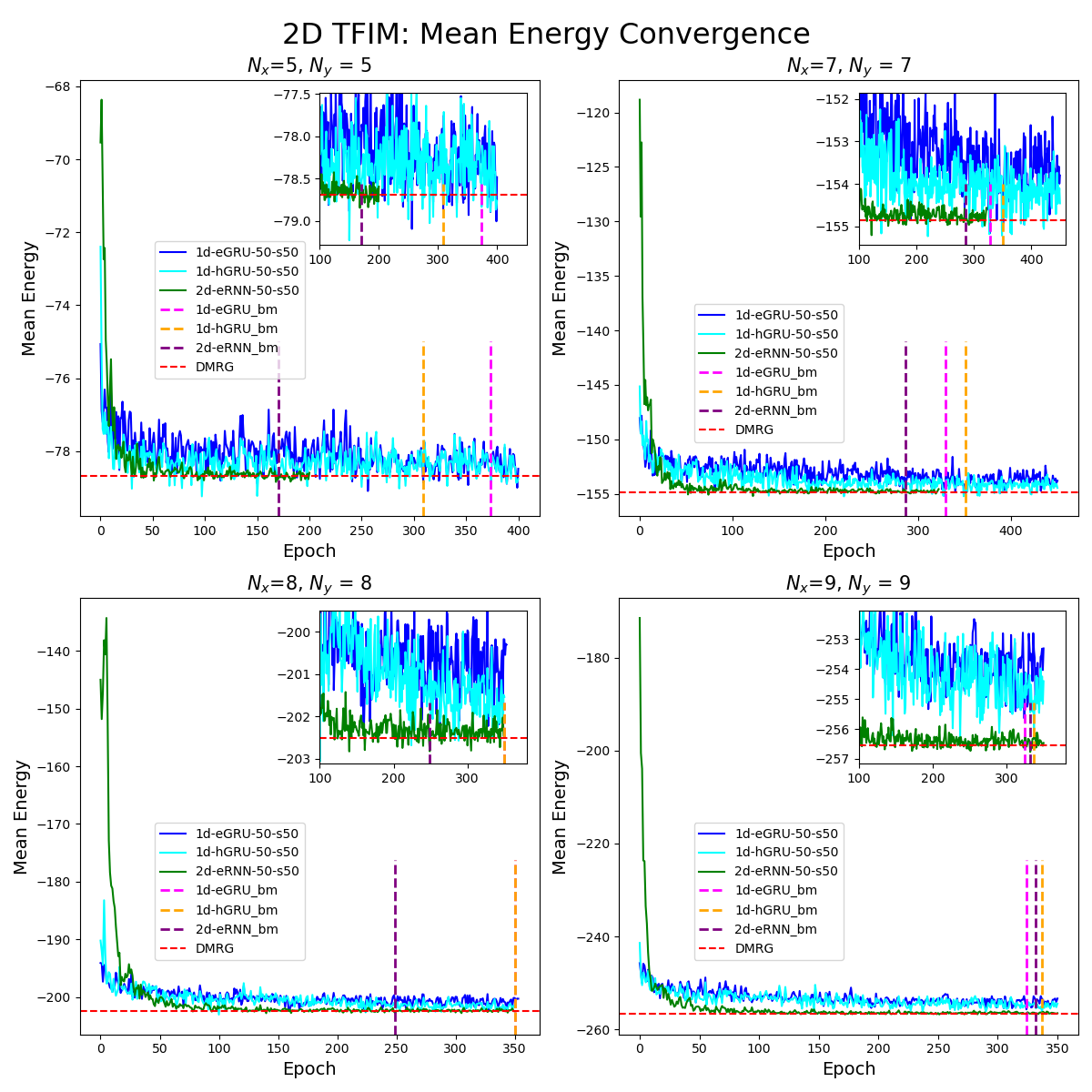}
\caption{Convergence curves of the mean energy using 2D Euclidean RNN and 1D Euclidean/hyperbolic GRU ansatzes for the 2D TFIM with ($N_x, N_y$) = (5,5), (7,7), (9,9), (8,8) lattices (clockwise from top). Each subfigure contains a zoomed-in plot showing a narrower range of energy values close to the exact DMRG value. The vertical dashed lines (whose legend labels contain the letters `\texttt{bm}') indicate the epochs at which the best models were saved. The epochs at which the best models were saved were automatically determined by the training loop such that after convergence is achieved, the best models not only have the lowest mean energy but also with the variance below the tolerance threshold (see the discussion in Section \ref{sec-res} just before Section 5.1). For $(N_x,N_y) = (5,5)$, the number of training epochs of \texttt{2d-eRNN} was much smaller than those of \texttt{1d-eGRU} and \texttt{1d-hGRU}, so the purple dashed line corresponding to the epoch at which the \texttt{2d-eRNN} best model was saved appear much earlier those corresponding to \texttt{1d-eGRU} and \texttt{1d-hGRU}.
When $(N_x,N_y) = (8,8)$, the best models for \texttt{1d-eGRU} and \texttt{1d-hGRU} were both saved the same epoch, thus only one vertical dashed line is visible, since these 2 dashed lines corresponding to  \texttt{1d-eGRU} and \texttt{1d-hGRU} coincide with each other. }\label{2d-tfim-meanE}
\end{figure}

\FloatBarrier
\clearpage
\subsection{1D Heisenberg $J_1J_2$ model}\label{sec-app-j1j2}

\subsubsection{Results without gradient clipping}\label{sec-app-j1j2-ngc}
In this section, we recorded the results obtained from training the models without using gradient clipping. 
\begin{table}[!h]
\centering
\begin{tabular}{lccccc}
\hline\hline
Ansatz & $J_2=0.0$ & $J_2=0.2$& $J_2=0.5$& $J_2=0.8$ &\\
\hline\hline
Euclidean GRU
&   \textbf{-21.2469} &    -17.6566   & -17.3347  &  -19.3759 &\\
&(0.0094) &   (0.0188) & (0.0128) & (0.0133) &
\\
Hyperbolic GRU
&  -20.4145 & \textbf{-19.7781} &   \textbf{-18.6643} &   \textbf{-20.0535} &\\
& (0.0112) &   (0.0113) &  (0.0066) &  (0.0120) &\\
\hline
Exact (DMRG)   & -21.9721& -20.3150 & -18.7500&  -20.9842 &\\
\hline
\end{tabular}
\caption{VMC inference results  (using $10^4$ samples) for the 1D $J_1J_2$ system with the Hamiltonian Eq.(\ref{Hj1j2}) with $N=50$, $J_1 = 1$ and $J_2=0.0,0.2, 0.5, 0.8$. For each ansatz, the mean $E$ is listed first, together with the standard error (in brackets). The best result at each $J_2$ value is noted in bold. These results were obtained by training the NQS without gradient clipping.} \label{tab:1dj1j2_res-old}
\end{table}
\begin{figure}[!h]
\centering
\includegraphics[width = .6\textwidth]{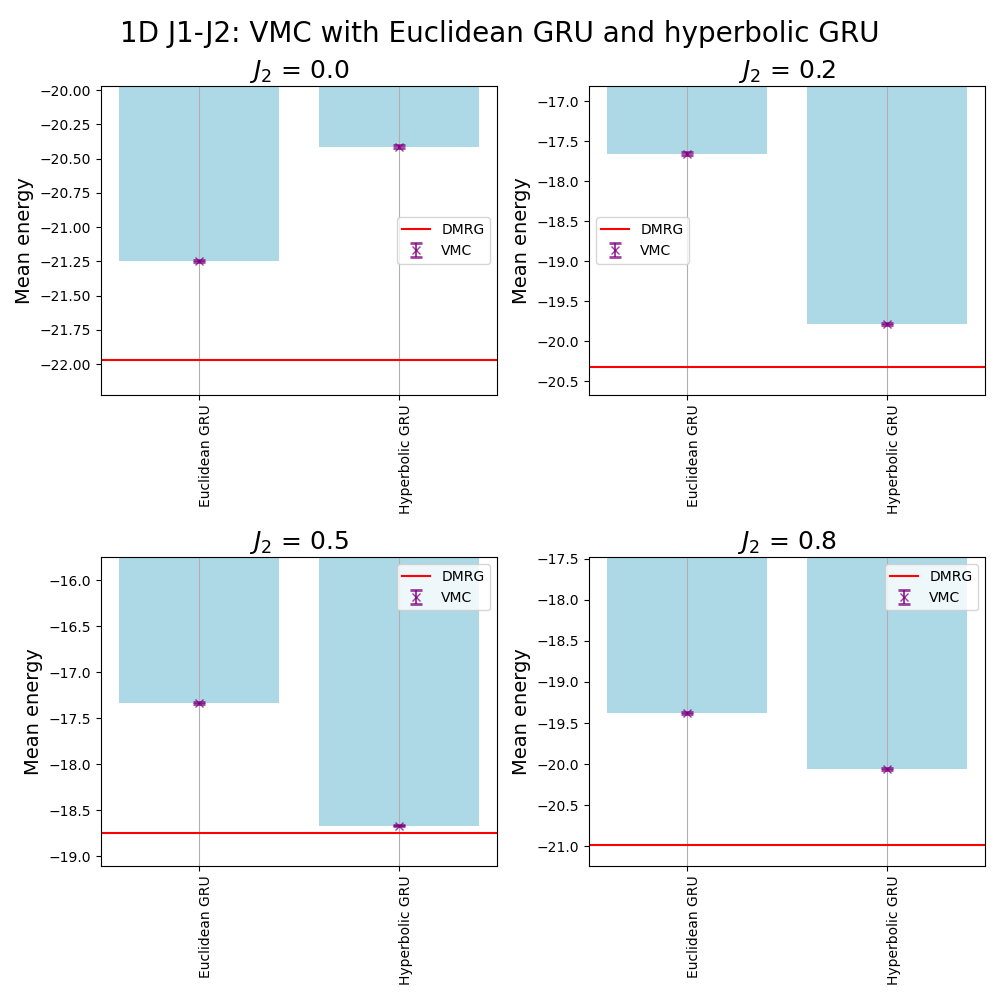}
\caption{Comparisons of the performances of Euclidean and hyperbolic GRU NQS ansatzes (trained without gradient clipping) listed in Table \ref{tab:1dj1j2_setting} from VMC runs of 1D $J_1J_2$ model with different $J_2$ values (clockwise from top: $J_2$ =0.0, 0.2, 0.8, 0.5) for $N=50$ spins. In each subfigure, the mean energy for each NQS ansatz is shown as a dot with an error bar (representing the standard error). The specific variants of Euclidean/hyperbolic GRU used in each of the four cases are listed in Table \ref{tab:1dj1j2_setting}. In this plot, we do not specify the exact variants, but label them simply as Euclidean GRU or hyperbolic GRU to emphasize the geometry of the ansatzes.}\label{1d-j1j2-n50-comparison-old}
\end{figure}
\newpage
\subsubsection{Convergence curves}
\begin{figure}[!h]
\centering
\includegraphics[width = .5\textwidth]{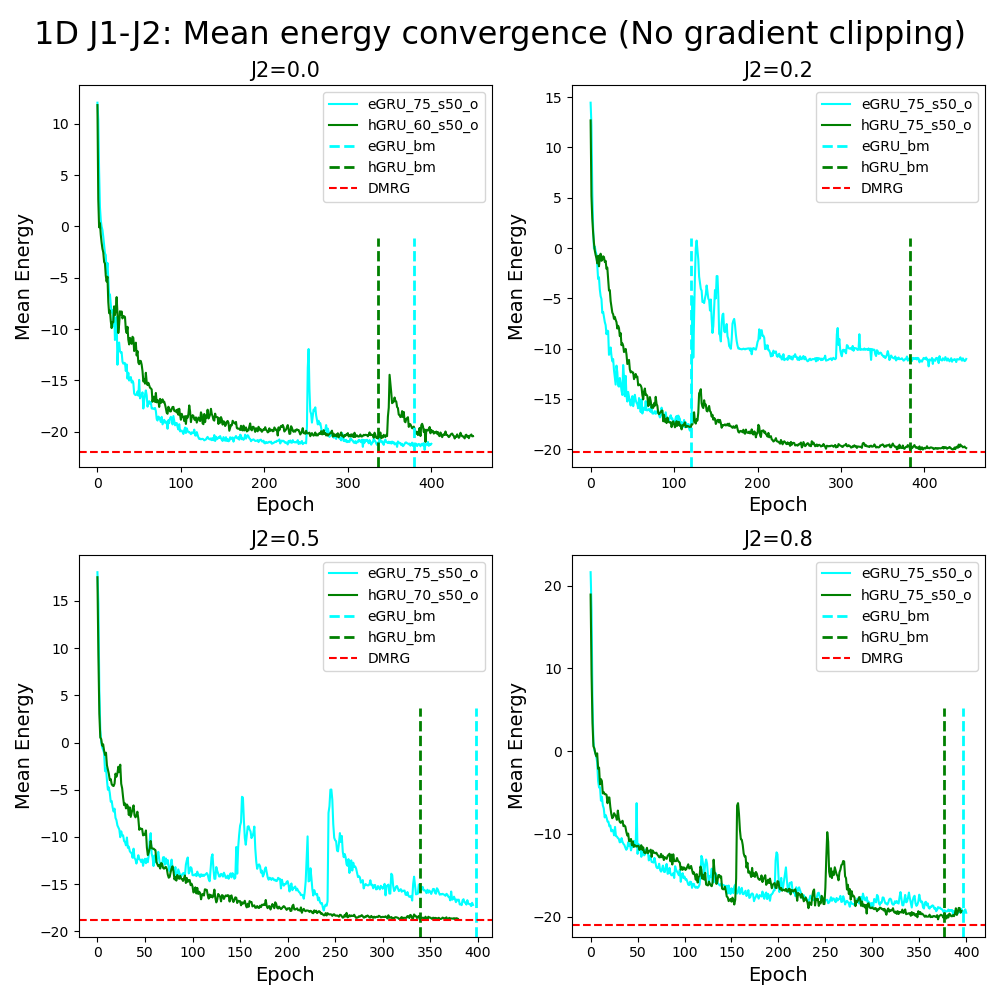}
\caption{Convergence curves of the mean energy using  Euclidean GRU and hyperbolic GRU from VMC runs of 1D Heisenberg $J_1J_2$ model with different $J_2$ values. These curves were obtained from training with no gradient clipping. Clockwise from top, $J_2$ = 0.0, 0.2, 0.8, 0.5 - for $N=50$ spins.
The vertical dashed lines (whose legend labels contain the letters `\texttt{bm}') indicate the epochs at which the best models were saved. The epochs at which the best models were saved were automatically determined by the training loop such that after convergence is achieved, the best models not only have the lowest mean energy but also with the variance below the tolerance threshold (see the discussion in Section \ref{sec-res} just before Section 5.1).
In this case, there are various kinks or jumps in the curves, indicating instabilities during the training process, espcially those corresponding to the Euclidean network. When $J_2 = 0.2$, the best model was saved right before there was a large jump in the energy.}\label{1d-j1j2-n50}
\end{figure}


\begin{figure}[!h]
\centering
\includegraphics[width = .55\textwidth]{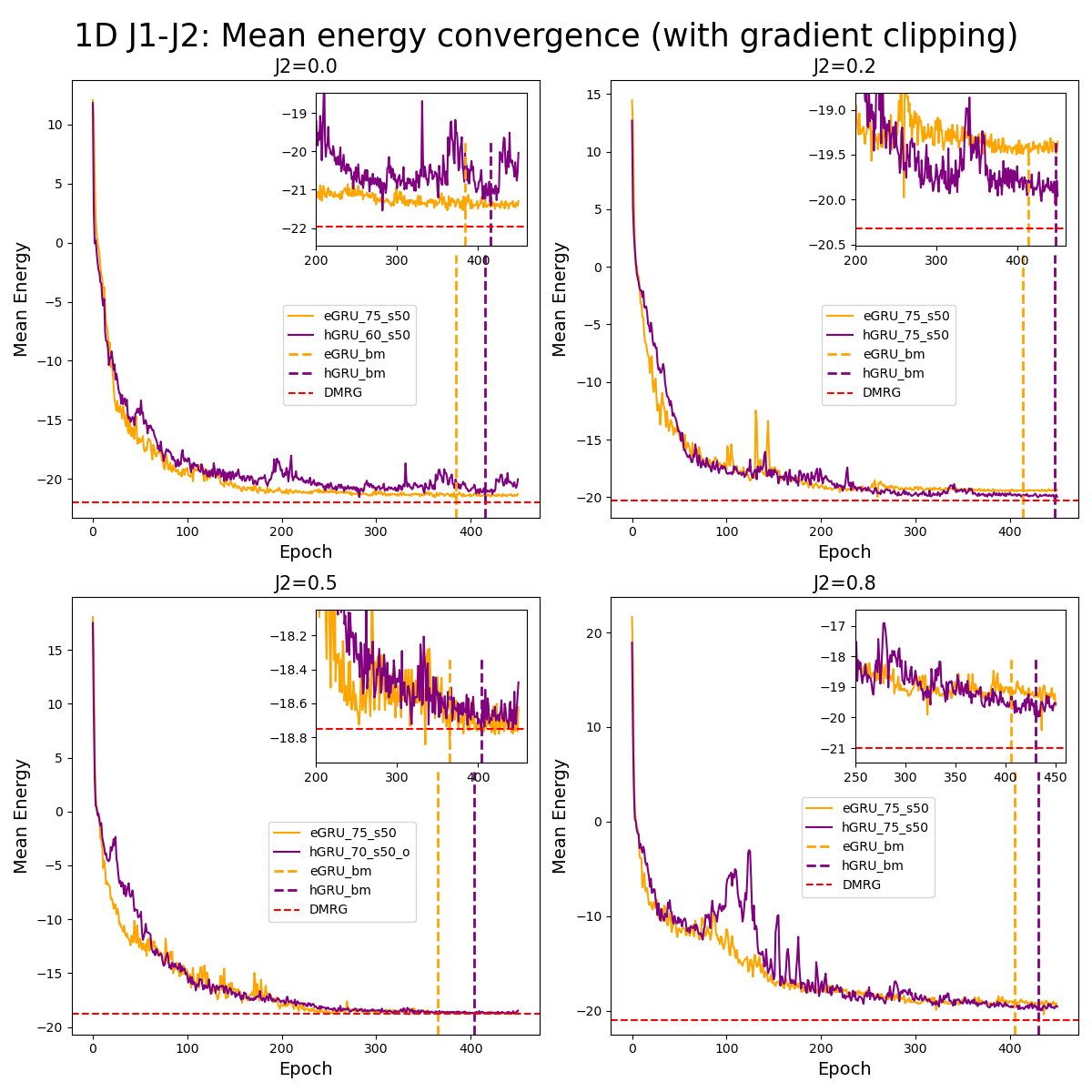}
\caption{Convergence curves of the mean energy using  Euclidean GRU and hyperbolic GRU from VMC runs of 1D Heisenberg $J_1J_2$ model with different $J_2$ values, \textit{with gradient clipping during the training process}, - clockwise from top, $J_2$ = 0.0, 0.2, 0.8, 0.5 - for $N=50$ spins. The vertical dashed lines (whose legend labels contain the letters `\texttt{bm}') indicate the epochs at which the best models were saved. The epochs at which the best models were saved were automatically determined by the training loop such that after convergence is achieved, the best models not only have the lowest mean energy but also with the variance below the tolerance threshold (see the discussion in Section \ref{sec-res} just before Section 5.1).}\label{1d-j1j2-n50-gc}
\end{figure}

\FloatBarrier
\clearpage

\begin{figure}[!h]
\centering
\includegraphics[width = .55\textwidth]{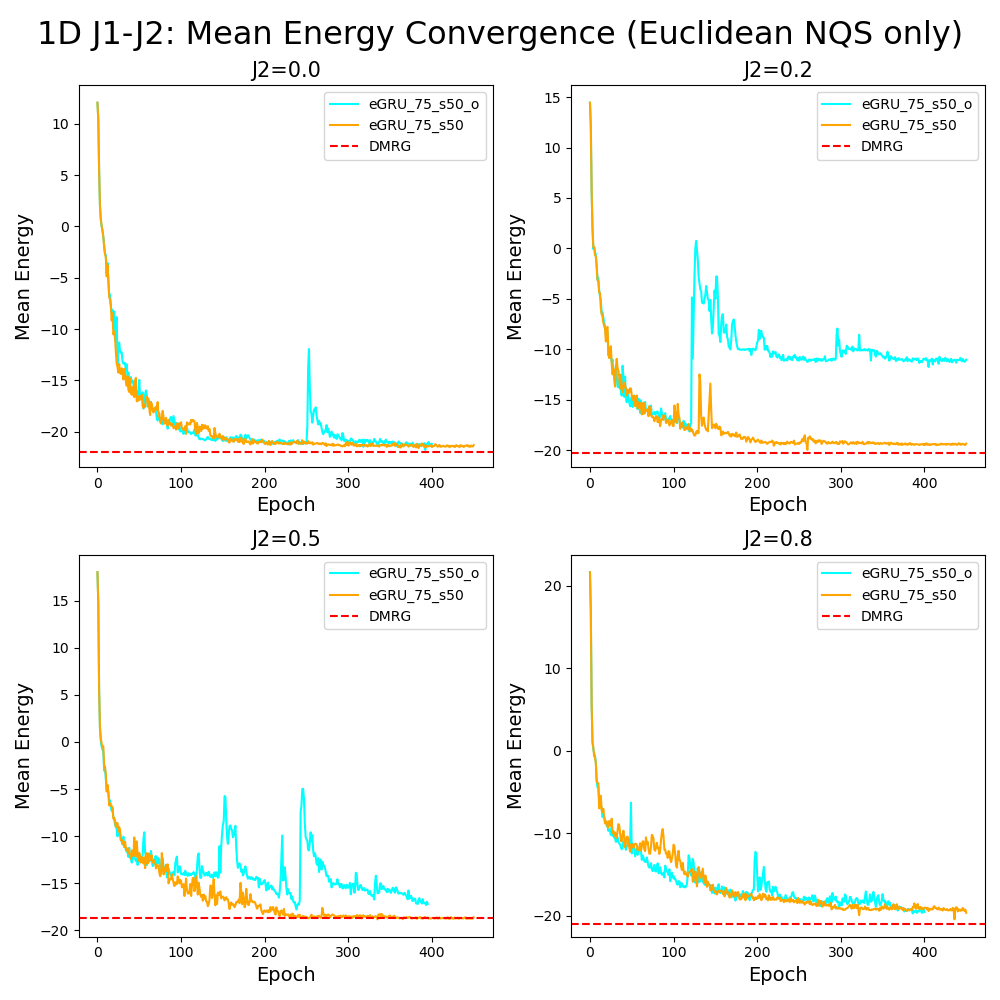}
\caption{Convergence curves of the mean energy of Euclidean GRU \textit{only} from VMC runs of 1D Heisenberg $J_1J_2$ model with different $J_2$ values \textit{with and without gradient clipping during the training process}. Curves obtained from training with no gradient clipping are marked by `-o' in their names, while those obtained from training with gradient clipping are not. Clockwise from top, $J_2$ = 0.0, 0.2, 0.8, 0.5 - for $N=50$ spins.}\label{1d-j1j2-n50-ernn-gc-ngc}
\end{figure}
\begin{figure}[!h]
\centering
\includegraphics[width = .5\textwidth]{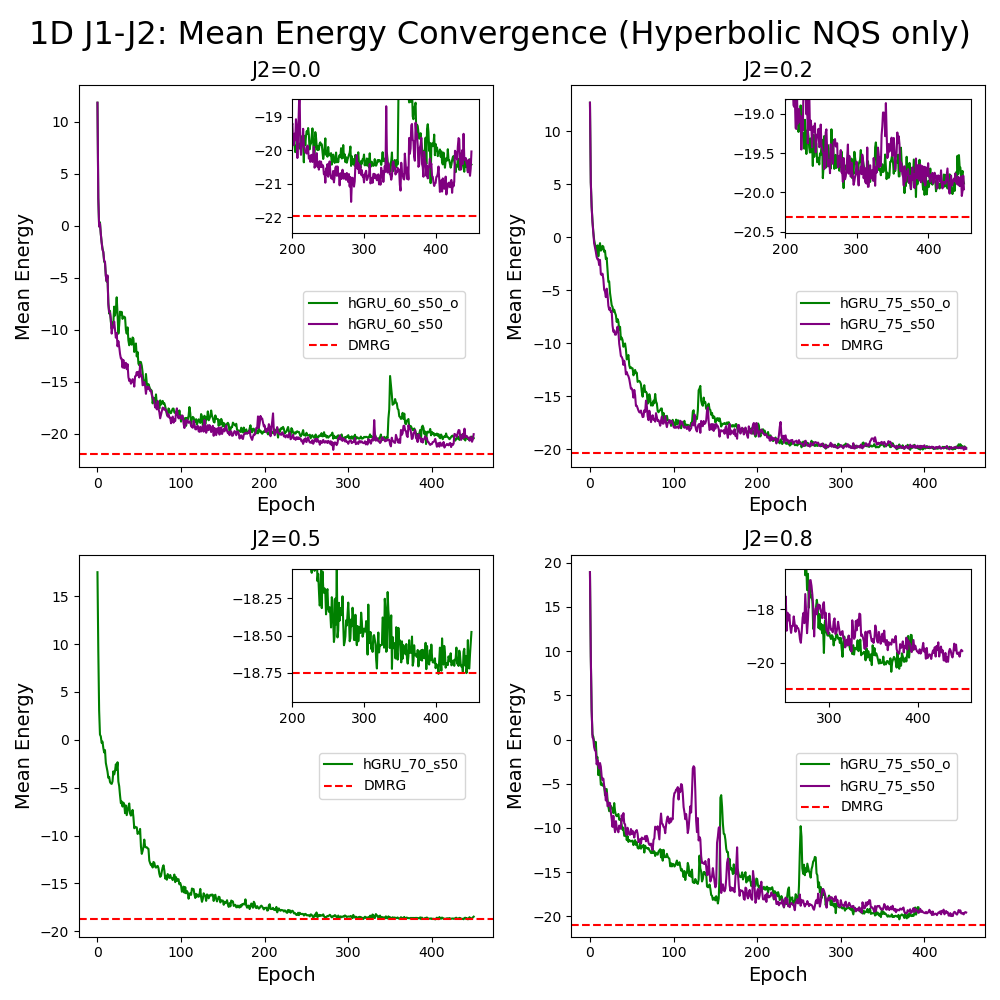}
\caption{Convergence curves of the mean energy of hyperbolic GRU \textit{only} from VMC runs of 1D Heisenberg $J_1J_2$ model with different $J_2$ values \textit{with and without gradient clipping during the training process}. Curves obtained from training with no gradient clipping are marked by `-o' in their names, while those obtained from training with gradient clipping are not. For the case of $J_2 = 0.5$, it was not necessary to perform the experiment again using gradient clipping due to no instability present in the training. Clockwise from top, $J_2$ = 0.0, 0.2, 0.8, 0.5 - for $N=50$ spins.}\label{1d-j1j2-n50-hrnn-gc-ngc}
\end{figure}
\FloatBarrier
\clearpage
\subsection{1D Heisenberg $J_1J_2J_3$ model} \label{sec-app-j1j2j3}
\begin{figure}[!h]
\centering
\includegraphics[width = .56\textwidth]{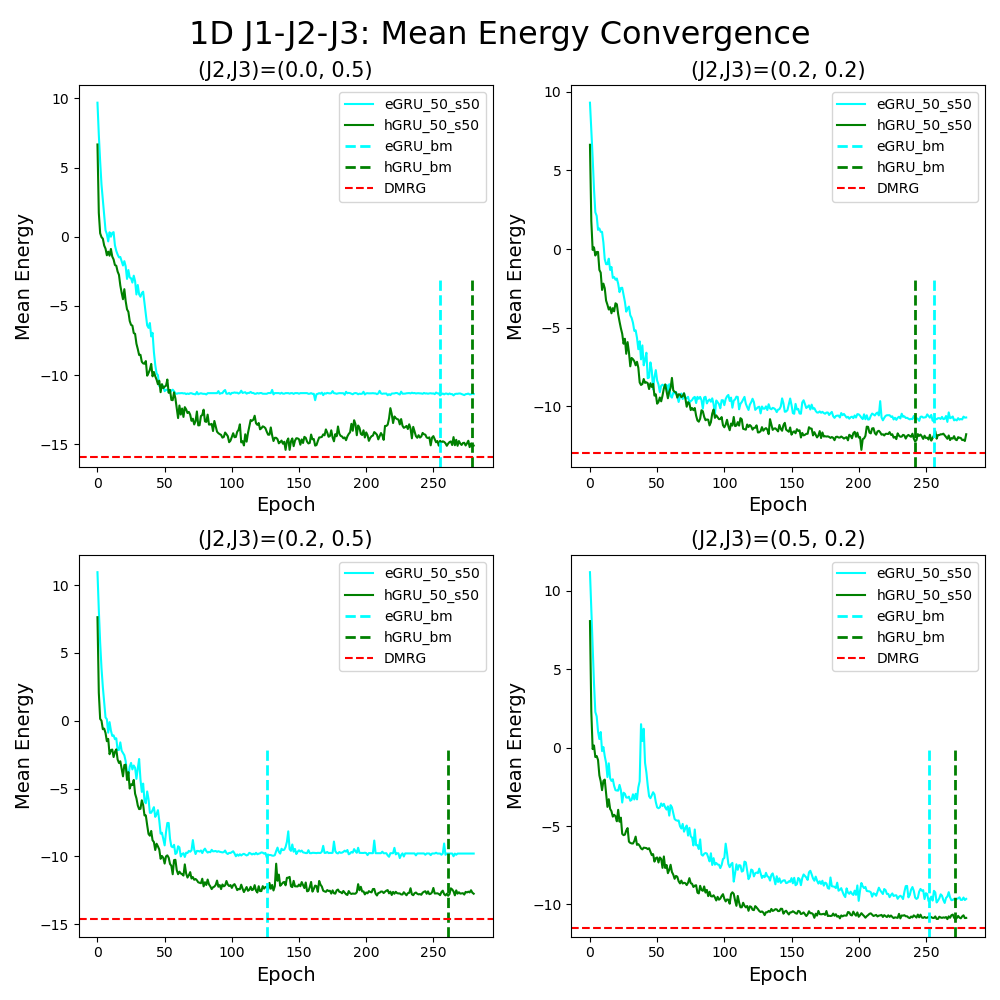}
\caption{Convergence curves of the mean energy using  Euclidean GRU and hyperbolic GRU from VMC runs of 1D Heisenberg $J_1J_2J_3$ model with different $(J_2,J_3)$ values - clockwise from top, ($J_2, J_3$) = (0.0,0.5), (0.2,0.2), (0.5,0.2), (0.2,0.5) -  for $N=30$ spins. The vertical dashed lines (whose legend labels contain the letters `\texttt{bm}') indicate the epochs at which the best models were saved. The epochs at which the best models were saved were automatically determined by the training loop such that after convergence is achieved, the best models not only have the lowest mean energy but also with the variance below the tolerance threshold (see the discussion in Section \ref{sec-res} just before Section 5.1).}\label{1d-j1j2j3-n30}
\end{figure}
\FloatBarrier

\begin{figure}[!h]
\centering
\includegraphics[width = .56\textwidth]{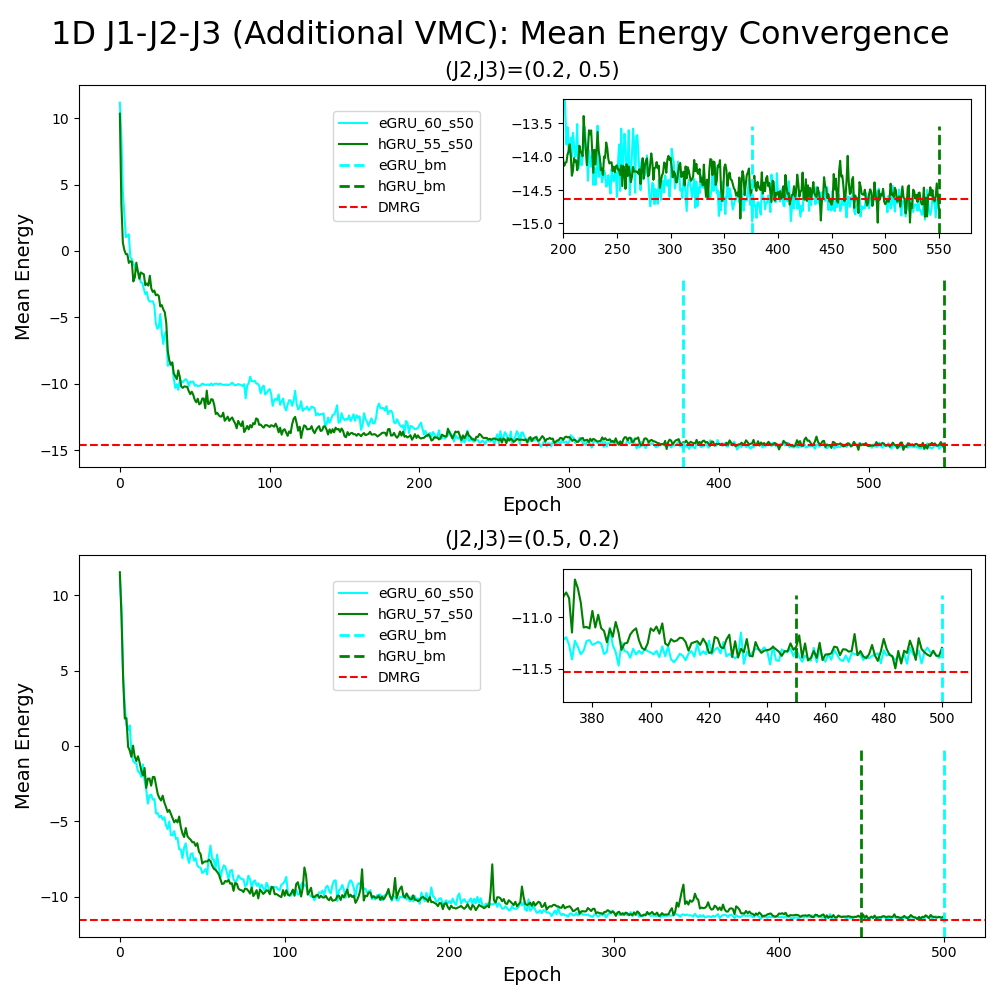}
\caption{Convergence curves of the mean energy using  Euclidean GRU and hyperbolic GRU from VMC runs of 1D Heisenberg $J_1J_2J_3$ model with different $(J_2,J_3)$ values - from top to bottom, ($J_2, J_3$) = (0.2,0.5), (0.5,0.2) -  for $N=30$ spins. Each subfigure contains a zoomed-in plot showing a narrower range of energy values close to the exact energy line. The vertical dashed lines (whose legend labels contain the letters `\texttt{bm}') indicate the epochs at which the best models were saved. The epochs at which the best models were saved were automatically determined by the training loop such that after convergence is achieved, the best models not only have the lowest mean energy but also with the variance below the tolerance threshold (see the discussion in Section \ref{sec-res} just before Section 5.1).}\label{1d-j1j2j3-n30_add}
\end{figure}

\FloatBarrier
\clearpage
\section{Data Availability Statement}
The data and results (in the form of Python scripts, Jupyter notebooks, and trained Euclidean/hyperbolic neural network models)  supporting the findings of this work are available at the following Github repository: \href{https://github.com/lorrespz/nqs_hyperbolic_rnn}{https://github.com/lorrespz/nqs\_hyperbolic\_rnn}


\end{document}